\documentclass[prd,aps,twocolumn,groupedaddress,showpacs]{revtex4}
\begin{document}
\input{psfig.sty}

\title{Towards a Realistic Neutron Star Binary Inspiral: Initial Data and
Multiple Orbit Evolution in Full General Relativity}

\author{Mark Miller}
\affiliation{238-332 Jet Propulsion Laboratory, 4800 Oak Grove Drive,
Pasadena, CA  91109}
\affiliation{McDonnell Center for the Space Sciences, Department of Physics,
Washington University, St. Louis, Missouri, 63130}

\author{Philip Gressman}
\affiliation{Mathematics Department, Princeton University, Princeton, NJ 08544}
\affiliation{McDonnell Center for the Space Sciences, Department of Physics,
Washington University, St. Louis, Missouri, 63130}

\author{Wai-Mo Suen}
\affiliation{McDonnell Center for the Space Sciences, Department of Physics,
Washington University, St. Louis, Missouri, 63130}
\affiliation{Physics Department, Chinese
University of Hong Kong, Shatin, Hong Kong}

\date{\today}

\begin{abstract}                                                        %
 
This paper reports on our effort in modeling realistic
astrophysical neutron star binaries in general relativity.
We analyze under what conditions the conformally flat quasiequilibrium 
(CFQE) approach can generate ``astrophysically relevant'' initial data,
by developing an analysis that
determines the violation of the CFQE approximation in the evolution of
the binary described by the full Einstein theory.  We
show that
the CFQE assumptions significantly violate the Einstein field equations
for corotating neutron stars
at orbital separations nearly double that of the 
innermost stable circular orbit 
(ISCO) separation, thus calling into question the 
astrophysical relevance of the
ISCO determined in the CFQE approach.

With the need to start numerical simulations at large orbital 
separation in mind, we push for stable and long term integrations 
of the full Einstein equations for the binary neutron star system.  
We demonstrate the stability 
of our numerical treatment and analyze the stringent requirements 
on resolution and size of the computational domain for an accurate 
simulation of the system.  
 
\end{abstract}


\maketitle

\section{Introduction}                                                  %
\label{sec:introduction}                                                %

The analysis of general relativistic binary neutron star processes
is an important, yet challenging, endeavor.  
The importance of understanding these
processes is rooted
in observational astronomy, both in gravitational wave astronomy
and high-energy electromagnetic wave astronomy.  Neutron star binaries
could be the central engines for some classes of
gamma-ray bursts, and they are definitely strong candidates as sources of
gravitational radiation detectable by
the up and coming generation of gravitational wave detectors
such as LIGO, VIRGO, TAMA, GEO, and LISA.
The challenge of understanding binary neutron star processes is rooted in
the complexity of both the nonlinear Einstein field equations and the
physical properties of the super-nuclear density matter which make up
the neutron stars.

Due to the complexity of the binary neutron star system,
various
levels of approximation have been 
employed (with various levels of success) as aids in understanding 
the details of the different stages of the inspiral of comparable 
mass binary neutron stars, from the quasistationary
inspiral stage through to the plunge and merger of the binary stars
to the ring-down of the final merged object.  These 
range from the approximation of the structure
of the neutron stars themselves (e.g., from point particle 
to finite sized perfect fluid models and the equations of 
state with different physical approximations) to the approximation
of general relativistic effects (e.g., from Newtonian
gravity, to post-Newtonian, to full 
general relativity).
Based on 
insights from point particle mechanics in general relativity
and orbits of finite size bodies in Newtonian gravity, one expects
that the early part of the inspiral process will be quasistationary, with
the secular (i.e. on timescales larger than the orbital timescale)
shrinkage of the orbit driven by gravitational radiation.
When the orbit shrinks to a small enough radius (but before touching),
there may or may not exist an innermost stable orbit (ISO) beyond
which dynamical processes drive the quasistationary inspiral into a
plunge phase.  Whether there is, in fact, a ``phase change'' 
from quasistationary inspiral to a plunge phase in the fully
relativistic theory, and whether or not this happens before any
other (hydrodynamical) dynamical instabilities, is an 
unsolved problem in full general relativity (see, e.g.,~\cite{Rasio92}
for answers in the Newtonian case).

A recent approach for the investigation of the later part of the
neutron star inspiral which has been studied in 
detail~\cite{Baumgarte98b,Baumgarte98c,Bonazzola97,Gourgoulhon01,
Marronetti:1999ya,Shibata98,Teukolsky98,Yo01,Duez00,Duez02} has drawn
much attention.  This treatment, which we refer to as the
conformally flat quasiequilibrium (CFQE) approach, is a procedure
for constructing general relativistic configurations that
correspond to compact, equal mass binary neutron stars 
in a quasiequilibrium,
circular orbit.  These individual configurations,
which we refer to as ``CFQE configurations'', are by themselves
solutions to the constraints of general relativity in 3+1 form,
i.e., the Hamiltonian and momentum constraints.
For a given equation
of state, each equal mass binary CFQE configuration
can be characterized by two parameters:
the separation of the two neutron stars and
the baryonic (rest) mass of each of the neutron stars 
(we only consider equal mass, corotating binary configurations here).
Taking the CFQE approximation one step further, one can construct 
an entire 4-dimensional spacetime 
by ``gluing'' these configurations together
as a time sequence of different CFQE configurations.  This construction,
which we will refer to as a CFQE-sequence, produces a spacetime
that solves 5 of the Einstein field equations (4 constraint equations
and trace of the extrinsic curvature 
equation)~\cite{Baumgarte98b,Gourgoulhon01}.
Since the total 
baryonic mass will be constant during the inspiral,
the CFQE approach takes constant baryonic CFQE-sequences as
a representation of the evolutionary sequence of the binary 
neutron star inspiral process;
secular evolution
of the orbit due to gravitational radiation brings one equilibrium
configuration into another with the same rest mass, forming the
CFQE-sequence~\cite{Baumgarte98b,Gourgoulhon01,
Duez00,Duez02}.

In this CFQE-sequence approach, one finds that for a corotational 
neutron star binary
system (which is the focus of this paper),
the ADM mass of each of the individual CFQE configurations
decreases along a constant rest mass
CFQE-sequence as the separation between the binary stars
decreases
until a minimum is attained, after which the ADM mass
increases as the separation decreases further. 
Using turning point theorems for exact equilibrium configurations
in general relativity~\cite{Sorkin82,Friedman88,Baumgarte98c},
this minimum point signals a secular instability in the
evolutionary sequence, and is commonly referred to as the
innermost stable circular orbit (ISCO) configuration.
In the Newtonian case, it has been shown that this secular instability
point is encountered in the evolutionary process {\it before} 
any dynamical instability 
is reached~\cite{Lai93}.

While it is reasonable to assume that the CFQE approximation holds
reasonably well
for highly separated neutron stars, 
it is not clear at which point along the evolutionary sequence the
CFQE approximation breaks down.  It is certainly not 
clear whether or not
the CFQE-sequence approximation is good for all neutron star
separations larger than the ISCO separation, due to the fact that one main
assumption in the turning-point 
theorems~\cite{Sorkin82,Friedman88,Baumgarte98c} used to interpret these
CFQE sequences is the assumption of exact equilibrium.  While CFQE
configurations are very close to equilibrium for large neutron star
separations, they become less so as the separation decreases.  Unfortunately,
it is exactly the small separation regime, i.e. near the ISCO configuration,
in which these theorems are being applied.  Also, the CFQE-sequence
approximation to full general relativity does not 
provide an
estimate of the error of its solutions, unlike, e.g.,  the
post-Newtonian approach where one can compute the next order post-Newtonian
terms to estimate the error of any post-Newtonian approximation.

The fact that each CFQE configuration satisfies the constraints of
general relativity suggests the use of these configurations as
initial data to full general relativistic calculations.
The setting of initial data is obviously an important issue in numerical
general relativistic astrophysical
simulations.  While all initial data configurations satisfying the constraints
of general relativity
(i.e., the Hamiltonian and momentum constraints)
are in principle legitimate initial data sets, they may not
be acceptable for the study of coalescence of astrophysical neutron
star binaries.  In order for the results of the numerical evolutions
to be relevant to observations, e.g., the gravitational waves emitted
in actual neutron star coalescences, we have to make sure that the initial data
actually corresponds to a configuration in an astrophysically
realistic inspiral.  It is not straightforward how one can
go about evaluating the usefulness
of the CFQE approach in approximating astrophysically relevant phenomena.
We note that while the CFQE approach leads to solutions of the
constraint equations, the CFQE-sequence approach is {\it not}
consistent with the full set of Einstein
equations.  

This paper is divided into four main sections. In 
section~\ref{sec:formulation}, we describe our fully consistent
general relativistic hydrodynamics code used in this paper.  We 
describe in detail the 3+1 formulation of the Einstein field equations
that we couple to the relativistic hydrodynamics equations, along
with details of the gauge conditions and discretization techniques used.

In section~\ref{sec:cfqe},
we describe the CFQE-sequence approximation, whose individual
configurations will be used as initial data for our fully consistent
general relativistic calculations.
We demonstrate that the CFQE-sequence definition of
the ISCO for corotating neutron star binaries may not be a relevant
concept by showing that if one
takes into account the spin energy of the neutron stars when
constructing the effective binding energy in the CFQE-sequence
approximation, then there is no longer a minimum in the
effective binding energy.

In section~\ref{sec:short_evolve},
we analyze
the key assumptions of the CFQE-sequence approximation by comparing
them with fully general relativistic simulations
using CFQE configurations as initial data.
We focus on the question of how well the CFQE-sequence
approximation for corotating neutron stars approximates full 
general relativity. 
This is done by comparing the assumptions in
building the CFQE-sequence to fully consistent
general relativistic calculations
using CFQE configurations as initial data to our numerical evolution
code.  Specifically, we
analyze the 
conformal flatness assumption and the
assumption of the existence of a Killing 
vector field.  We devise a number of invariant measures of these
assumptions, and monitor them in our fully consistent, 
general relativistic simulations.
We find that, as expected, the accuracy of the CFQE-sequence approximation
increases for increasing neutron star separation.  
We present a general algorithm for evaluating whether or not any
CFQE configuration can be thought of as astrophysically realistic initial
data by analyzing how much the CFQE-sequence approximation violates
the Einstein field equations.
We demonstrate this method by showing that, for any given tolerance,
one can find a CFQE configuration whose subsequent evolution
in full general relativity will not violate the Einstein field 
equations within a small fraction of an orbit.

In section~\ref{sec:long_evolve}, we analyze
the long timescale (i.e. multiple
orbits of the binary system) numerical evolutions of CFQE 
initial data configurations using our general relativistic hydrodynamics code.
In particular, we use a CFQE configuration which has a $15\%$ larger proper 
separation than that of the ISCO CFQE configuration.  We analyze the
orbital decay rate on multiple orbit timescales.
Using care in estimating
the numerical truncation error as well as the errors introduced by
placing the computational boundaries at a finite distance from the
neutron star binaries, we find that
while the computational resources
at our disposal are sufficient to get a reasonable handle on the
truncation error, 
the errors
introduced by the placement of the boundary of the computational domain 
can have a significant impact on the dynamics of the neutron star
inspiral.  We demonstrate that this is true even if the linear 
dimensions of the computational domain
are as large as one half
the size of the gravitational waves being emitted, which
is a fairly large computational domain by today's numerical relativity
standards.  
We conjecture that
presently available computational resources will not allow a unigrid 
finite difference code
to decrease the discretization parameter
$\Delta x$ sufficiently and simultaneously
increase the distance from the binary system
to the computational domain boundary sufficiently in order
to guarantee that the induced numerical errors will not significantly 
affect the details of the inspiral process.  Mesh refinement techniques
and/or better outer boundary conditions 
will be needed in order to accurately simulate the physics on 
the length scales of the compact object as well as the length scales of
the gravitational radiation. 

\section{Formulation and Discretization of the Einstein Equations
Coupled to a Perfect Fluid}
\label{sec:formulation}

Our code numerically solves the Einstein field equations coupled
to a relativistic perfect fluid.
The gravitational degrees of
freedom are geometrically encoded in the 4-metric, $g_{\mu \nu}$,
which are governed by the Einstein field equations
\begin{equation}
G_{\mu \nu} = 8 \pi T_{\mu \nu},
\label{eq:einstein}
\end{equation}
where $G_{\mu \nu}$ is the Einstein tensor and $T_{\mu \nu}$ is the
stress-energy tensor of the perfect fluid, given as
\begin{equation}
T_{\mu \nu} = \rho h u_{\mu} u_{\nu} + P g_{\mu \nu}.
\end{equation}
Here, we have set the gravitational constant $G$ and the speed of
light $c$ to be identically $1$.  The 4-velocity of the perfect
fluid is denoted as $u_{\mu}$, and $\rho$, $P$, and $h$ are the baryonic mass
density, pressure, and specific enthalpy, respectively, of the fluid.  
The equations of motion governing the perfect fluid are the conservation
of stress-energy and baryonic mass
\begin{eqnarray}
{\nabla}_{\mu} T^{\mu \nu} & = & 0 \nonumber \\
{\nabla}_{\mu} \left ( \rho u^{\mu} \right ) & = & 0.
\label{eq:hydro}
\end{eqnarray}
These represent five equations governing the five degrees of freedom
of the perfect fluid (the mass density, energy density, and velocity).
The entire system of equations is closed by choosing an equation of state
for the pressure $P$ as a function of the baryonic mass density
and internal energy density of the fluid.

\subsection{The Einstein equations in 3+1 form}

In order to numerically solve the Einstein field 
equations, Eq.~\ref{eq:einstein}, we must cast the equations as
an initial value problem.  In order to facilitate this, we introduce
a foliation of the spacetime into spacelike hypersurfaces 
where the coordinate $t$ labels each
spacial hypersurface.  We furthermore introduce Cartesian coordinates $x^i$
on each spacelike hypersurface.  The line element can now be
written as 
\begin{equation}
ds^2 = ({\beta}^2 - {\alpha}^2) dt^2 + 2 {\beta}_i \: dt \, dx^i 
   + {\gamma}_{ij} \: dx^i \, dx^j
\end{equation}
where the shift vector $\beta^i$ is a 3-vector defined on each spacelike
hypersurface, $\alpha$ is the lapse function, and ${\gamma}_{ij}$
is the 3-metric.  We denote ${\gamma}^{ij}$ as the inverse of the
3-metric ${\gamma}_{ij}$, such 
that ${\gamma}^{ij} {\gamma}_{jk} = \delta^i_k$.

There are many ways to formulate the Einstein equations in 
an initial value, 3+1 form.  The standard ``ADM'' 3+1
formulation~\cite{Arnowitt62} writes the six space-space
components of the Einstein 
equations, Eq.~\ref{eq:einstein}, as 12 equations that are first order in time
\begin{eqnarray}
{\cal L}_t \gamma_{ij} & = & - 2 \alpha K_{ij} + 
   {\cal L}_{\vec{\beta}} \gamma_{ij} \\
{\cal L}_t K_{ij} & = & \alpha \, {}^{(3)} \! R_{ij} - 
   2 \alpha {K_{i}}^k K_{jk} 
   + \alpha K K_{ij} - \\
 & & {\cal D}_i {\cal D}_j \alpha + 
   {\cal L}_{\vec{\beta}} K_{ij} - \alpha \, {}^{(4)} \! R_{ij},
\end{eqnarray}
where $K_{ij}$ is the extrinsic curvature of a spacelike hypersurface.
Here, ${\cal L}$ is the Lie derivative operator, ${\cal D}_i$ is the 
covariant derivative operator compatible with the 3-metric ${\gamma}_{ij}$,
$K$ is the trace of the extrinsic curvature,
${}^{(3)} \! R_{ij}$ is the 3-Ricci tensor, while 
${}^{(4)} \! R_{ij}$ are the components of the 4-Ricci tensor that 
represent the perfect fluid source terms to the Einstein equations.
The remaining four Einstein equations are the constraint equations,
which are analytically 
satisfied on each of the spacelike hypersurfaces as long as 
they are satisfied on the initial slice.  
The Hamiltonian and momentum constraints are
\begin{eqnarray}
\label{eq:ham_constraint}
{}^{(3)} \! R + K^2 - K_{ij} K^{ij} - 2 {\alpha}^2 G^{tt} = 0 \\
{\cal D}_j {K^j}_i - {\cal D}_i K - \alpha {G_i}^t = 0
\label{eq:mom_constraint}
\end{eqnarray}

While one could base a numerical evolution code on the ADM equations,
recent results (both empirical~\cite{Baumgarte99,Alcubierre99d} and
analytical studies~\cite{Alcubierre99e,Miller00,Sarbach02a}) indicate that
a more suitable choice would be the 
so-called BSSN
formulation~\cite{Shibata95,Baumgarte99,Alcubierre99d}.
One feature of this formulation
of the Einstein equations is that the 3-metric is decomposed 
into a conformal factor $\phi$ and a conformal 3-metric
$\tilde{\gamma}_{ij}$ as
\begin{equation}
\gamma_{ij} = e^{{4 \phi}} \tilde{\gamma}_{ij},
\end{equation}
where the determinant of the conformal 3-metric $\tilde{\gamma}_{ij}$ is
identically $1$.  Similarly, the extrinsic curvature is decomposed into
its trace and trace-free parts
\begin{equation}
K_{ij} = \frac {1}{3} \gamma_{ij} K + e^{{4 \phi}} \tilde{A}_{ij},
\end{equation}
where $\tilde{A}_{ij}$ is referred to as the conformal trace-free
extrinsic curvature, such that $\tilde{A}_{ij} \tilde{\gamma}^{ij} = 0$.
In addition to the decomposition of the traditional ADM variables,
a key ingredient in the BSSN formulation is the introduction
of three new evolved variables, namely the 
three conformal connection functions ${\tilde{\Gamma}}^i$ 
\begin{equation}
{\tilde{\Gamma}}^i = - {\partial}_j \tilde{\gamma}^{ij}.
\end{equation}
The final form of the evolution equations which we use in the numerical
evolution of the Einstein field equations is given as
\begin{eqnarray}
\label{eq:bssn_phi}
\frac {\partial \phi}{\partial t} & = &
   - \frac {1}{6} \alpha K + \frac {1}{6} {\cal D}_k \beta^k \\
\label{eq:bssn_trk}
\frac {\partial K}{\partial t} & = &
   \alpha \tilde{A}^{ij} \tilde{A}_{ij} + 
   \frac {1}{3} \alpha K^2 - 
   \gamma^{ij} {\cal D}_i {\cal D}_j \alpha + {\cal L}_{\vec{\beta}}K + 
   \nonumber \\
   & & 2 \alpha^3 G^{tt} -
    \alpha \, {}^{(4)} \! {R_i}^i \\
\frac {\partial \tilde{\gamma}_{ij}}{\partial t} & = &
   - 2 \alpha \tilde{A}_{ij} - \frac {2}{3} \tilde{\gamma}_{ij} 
   {\cal D}_k \beta^k +
   e^{-4 \phi} {\cal L}_{\vec{\beta}} \gamma_{ij} \\
\frac {\partial \tilde{A}_{ij}}{\partial t} & = &
   \alpha e^{-4 \phi} \, {}^{(3)} \! R_{ij} -
   \frac {1}{3} \alpha \tilde{\gamma}_{ij} (\tilde{A}_{kl} \tilde{A}^{kl} -
      \frac {2}{3} K^2 ) + \nonumber \\
   & & \alpha K \tilde{A}_{ij} - 2 \alpha \tilde{A}_{ik} {\tilde{A}_j}^k - 
   e^{-4 \phi} {\cal D}_i {\cal D}_j \alpha + \nonumber \\
   & & \frac {1}{3} \tilde{\gamma}_{ij} {\cal D}_k {\cal D}^k \alpha +
   e^{-4 \phi} {\cal L}_{\vec{\beta}}(e^{4 \phi} \tilde{A}_{ij}) - 
   \nonumber \\
   & & \frac {2}{3} \tilde{A}_{ij} {\cal D}_k {\beta}^k -
   \frac {2}{3} \alpha^3 \tilde{\gamma}_{ij} G^{tt} - \nonumber \\
   & & \alpha e^{-4 \phi} \, {}^{(4)} \! R_{ij} +
   \frac {1}{3} \alpha \tilde{\gamma}_{ij} \, {}^{(4)} \! {R_k}^k \\
\frac {\partial \tilde{\Gamma}^{i}}{\partial t} & = &
   - 2 \tilde{A}^{ij} {\cal D}_j \alpha -
   \frac {4}{3} \alpha \tilde{\gamma}^{ij} {\cal D}_j K +
   12 \alpha \tilde{A}^{ij} {\cal D}_j \phi  - \nonumber \\
   & & {\partial}_j \left ( e^{4 \phi} {\cal L}_{\vec{\beta}} \gamma^{ij} +
      \frac {1}{3} \tilde{\gamma}^{ij} \gamma^{kl} {\cal L}_{\vec{\beta}}
         \gamma_{kl} \right ) + \nonumber \\
   & & 2 \alpha {\tilde{\Gamma}^i}_{jk} \tilde{A}^{jk} -
   2 \alpha^2 e^{4 \phi} G^{it}.
\label{eq:bssn_gamma}
\end{eqnarray}

\subsection{general relativistic hydrodynamics}

We also perform a 3+1 decomposition of the hydrodynamics
equations, Eq.~\ref{eq:hydro}.
Note that
the 4-velocity $u_{\mu}$ is
normalized $u^{\mu} u_{\mu} = -1$, so that its components
can be written in terms of the three
spatial velocity components $v^i$ as
\begin{equation}
\{u^\mu\} = \frac {W}{\alpha} \{ 1,\alpha v^i - \beta^i\},
\end{equation}
where $W$ is the Lorentz factor $W=1/\sqrt{1 - \gamma_{ij} v^i v^j}$.
The specific enthalpy, $h$, is given
as
\begin{equation}
h = 1 + \epsilon + P/\rho,
\end{equation}
where $\epsilon$ is the specific internal energy density.

The general relativistic hydrodynamics equations, Eq.~\ref{eq:hydro},
can be written in first order, flux conservative form as
\begin{equation}
\partial_t \vec{\cal U} + \partial_i \vec{F}^i = \vec{S},
\label{eq:hydro_fosh}
\end{equation}
where the conservative hydrodynamical variables $\vec{\cal U}$ are written
in terms of the primitive variables $\{\rho,v^i,\epsilon\}$ as
\begin{equation}
\vec{\cal U} = \left [ \begin{array}{c} D \\ S_j \\ \tau \end{array} \right ] =
   \left [ \begin{array}{c} \sqrt{\gamma} W \rho \\
      \sqrt{\gamma} \rho h W^2 v_j \\
      \sqrt{\gamma}(\rho h W^2 - P - W\rho)
   \end{array} \right ].
\end{equation}
The flux vector $\vec{F}^i$ is written as
\begin{equation}
\vec{F}^i = \left [ \begin{array}{c}
   \alpha \left ( v^i - \beta^i/\alpha \right ) D \\
   \alpha \left ( (v^i - \beta^i/\alpha)S_j +
      \sqrt{\gamma} P \delta^i_j \right ) \\
   \alpha \left ( (v^i - \beta^i/\alpha) \tau +
      \sqrt{\gamma} v^i P \right ) \end{array} \right ],
\label{eq:hydro_flux}
\end{equation}
and the source vector $\vec{S}$ is written as
\begin{equation}
\vec{S} = \left [ \begin{array}{c}
   0 \\
   \alpha \sqrt{\gamma} T^{\mu \nu} g_{\nu \sigma} {\Gamma^\sigma}_{\mu j} \\
   \alpha \sqrt{\gamma} (T^{\mu t} \partial_\mu\alpha - \alpha
      T^{\mu \nu} {\Gamma^t}_{\mu \nu} ) 
   \end{array} \right ]
\end{equation}

\subsection{discretization techniques}
\label{sec:discrete}
We discretize each of the 3 spatial coordinate variables $\{x,y,z\}$ 
using a constant spacing $\{\Delta x,\Delta y,\Delta z\}$, e.g.,
\begin{equation}
x_i = x_0 + i \, \Delta x, \;\;\; i=0,\cdots,n_x - 1.
\end{equation}
We discretize the time coordinate $t$ as
\begin{equation}
t_{n+1} = t_n + \Delta t
\end{equation}
where we set $\Delta t = 0.25 \: \Delta x$ for 
all dynamical simulations performed
in this paper.

Due to the fundamental differences in the phenomena being
described by the Einstein field equations, Eq.~\ref{eq:einstein}, and the 
relativistic hydrodynamics equations, Eq.~\ref{eq:hydro}, the discretization
methods that we employ for the two sets of equations are 
very different.  In the case of the Einstein field equations,
we expect the dynamical degrees of
freedom to remain 
smooth and continuous.  In the case of the relativistic
hydrodynamical equations, we know that shocks (discontinuities) can
easily form in the physical degrees of freedom.  Thus, the 
discretization method used for the hydrodynamical equations will be
more complicated in order to allow for the accurate treatment of 
shock propagation.  The approach we use will be based on a finite
differencing scheme employing High Resolution Shock Capturing
(HRSC) techniques.  In order to use these techniques, a complete 
knowledge of the characteristic information is needed.  We therefore
require the eigenstructure of the Jacobian matrices in 
Eq.~\ref{eq:hydro_fosh}, 
namely $\partial \vec{F}^i/\partial \vec{{\cal U}}$ for the
flux vector $\vec{F}^i$ defined in Eq.~\ref{eq:hydro_flux}.  This
is not a straightforward task, since the flux $\vec{F}^i$ is
expressed as a function of both the primitive and evolved hydrodynamical
variables.  What we require are a complete set of eigenvectors
$[\vec{r}_i]$ and corresponding
eigenvalues ${\lambda}_i$ such that
\begin{equation}
\left [ \frac {\partial \vec{F^x}}{\partial \vec{\cal{U}}} \right ]
[\vec{r}_i] = {\lambda}_i [\vec{r}_i], \;\;\; i=1,\cdots,5.
\end{equation}
(here, we present the spectral decomposition for the $x$-component
of the Jacobian, since the decomposition for the other two spatial 
components of the Jacobian can be obtained by a straightforward
permutation of the spatial coordinates $\{x,y,z\}$).  It turns out that
the spectral decomposition contains a triply degenerate eigenvalue
\begin{equation}
{\lambda}_1 = {\lambda}_2 = {\lambda}_3 = \alpha v^x - {\beta}^x.
\end{equation}
A set of linearly independent vectors that
span this degenerate space is given by

\begin{widetext}

\begin{equation}
\vec{r}_1 = { \left[\frac {\kappa}{h W (\kappa - \rho {c_s}^2)},
               v_x,v_y,v_z,
               1 - \frac {\kappa}{h W (\kappa - \rho {c_s}^2)}\right]}^T,
\end{equation}
\begin{equation}
\vec{r}_2 = { \left[W v_y, h (\gamma_{xy} + 2 W^2 v_x v_y),
                      h (\gamma_{yy} + 2 W^2 v_y v_y),
                      h (\gamma_{yz} + 2 W^2 v_y v_z),
                      v_y W (2 W h - 1)\right]}^T,
\end{equation}
\begin{equation}
\vec{r}_3 = { \left[W v_z, h (\gamma_{xz} + 2 W^2 v_x v_z),
                      h (\gamma_{yz} + 2 W^2 v_y v_z),
                      h (\gamma_{zz} + 2 W^2 v_z v_z),
                      v_z W (2 W h - 1)\right]}^T.
\end{equation}
The other two eigenvalues are given by
\begin{equation}
{\lambda}_{\pm} = \frac {\alpha}{1 - v^2 {c_s}^2}
\left \{ v^x (1 - {c_s}^2) \pm \sqrt{{c_s}^2 (1 - v^2)
   \left [ \gamma^{xx} (1 - v^2 {c_s}^2) - v^x v^x (1 - {c_s}^2) \right ]}
\right \} - {\beta}^x,
\end{equation}
with corresponding eigenvectors
\begin{equation}
\vec{r}_{\pm} = { \left[1,
   h W \left ( v_x - \frac {v^x - ({\lambda}_{\pm} + {\beta}^x)/{\alpha}}
           {\gamma^{xx} - v^x ({\lambda}_{\pm} + {\beta}^x)/{\alpha}} \right ),
   h W v_y, h W v_z,
   \frac {h W (\gamma^{xx} - v^x v^x)}{\gamma^{xx} - v^x
           ({\lambda}_{\pm} + {\beta}^x)/{\alpha}} - 1 \right] }^T,
\end{equation}
\end{widetext}
where the relativistic speed of sound
in the fluid $c_{s}$ is given by (see, e.g.,~\cite{Landau75})
\begin{equation}
c_{s}^2 =  { \left. {\frac {\partial P}{\partial E}} \right |}_{\cal{S}} =
\frac {\chi}{h} + \frac {P}{ {\rho}^2} \frac {\kappa}{h}.
\end{equation}
We have set 
$\chi = {\left.\frac {\partial P}{\partial \rho} \right |}_\epsilon$
and
$\kappa = {\left. \frac {\partial P}{\partial \epsilon} \right |}_\rho$.
$\cal{S}$ is the entropy per particle and
$E$ is the total rest energy density which in our case is
$E = \rho + \rho \epsilon$.  We use the above characteristic information
to calculate the numerical fluxes $(\vec{f}^{*})^x$ using the 
Piecewise-Parabolic Method (PPM), described in~\cite{Woodward84,Colella84}.
While the PPM method has been extended to special relativistic
applications (see, e.g.,~\cite{Marti96}), this is the first fully general
relativistic application of the method.  The 
discretization of the flux terms in Eq.~\ref{eq:hydro_fosh}
are written
\begin{equation}
\frac {\partial \vec{F}^x}{\partial x} = \frac {{(\vec{f}^{*})}_{i + 1/2} -
{(\vec{f}^{*})}_{i - 1/2}}{\Delta x} + {\cal O}( {\Delta x}^2).
\end{equation}
In order to update the discretized hydrodynamical variables, we simply 
perform a two-step predictor-corrector method, in order that the 
entire hydrodynamical update is done in a fully second order manner
in {\it both} space and time (modulo the points where the
hydrodynamical variables obtain local extrema, 
where the accuracy of the spatial
derivatives drop down to first order in space. This is a well known
property of the so-called Godunov schemes; see, e.g.,~\cite{Hirsch92}).

\begin{figure}
\vspace{0.0cm}
\hspace{0.0cm}
\psfig{figure=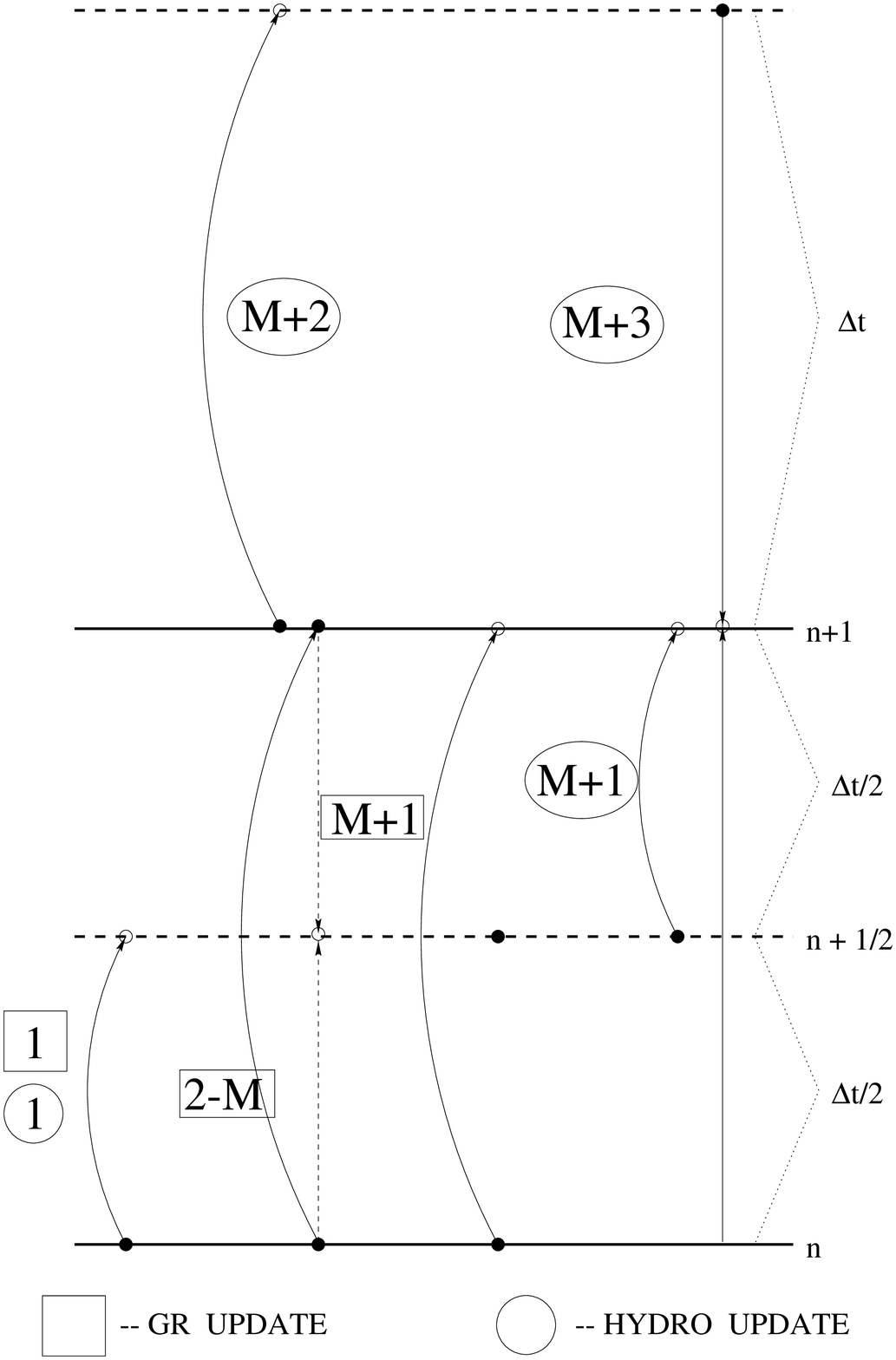,width=8cm}
\caption{A representation of 
the coupling between the
hydrodynamic predictor-corrector scheme (circles)
and the iterated Crank-Nicholson method
used for the integration of the Einstein field equations (squares).
{\bf STEP 1:} Simultaneous update of the general relativity and 
hydrodynamic equations
via a Euler-predictor step (first order in time) to the half timestep $n+1/2$.
{\bf STEP 2 through M:} Update of the general relativity 
equations via an iterative
Crank-Nicholson scheme (second order accurate in time) to the
$n+1$ timestep, then compute a corrected $n+1/2$ state by averaging
the $n+1$ and $n$ states.
{\bf STEP M+1: } Simultaneous update of the general relativity equations via
a leapfrog step (second order in time) based on the $n$ and
$n+1/2$ states, and the hydrodynamics equations via the second half of
the Euler-predictor step (first half applied in step 1)
using a method of lines.
{\bf STEP M+2: } Update of the hydrodynamic equations to a virtual
$n+2$ timestep via a (first order in time) Euler-corrector step
using method of lines.  {\bf STEP M+3: } A second order (in time)
hydrodynamics update is obtained by averaging the corrected quantities
of step $M+2$ at level $n+2$ and the original variables at level $n$.}
\vspace{0.0cm}
\label{fig:icn_couple}
\end{figure}

As previously stated, since the fields describing the gravitational
degrees of freedom are expected to remain smooth, we simply
perform centered-in-space discretizing of the spatial derivatives
in Eqs.~\ref{eq:bssn_phi}-~\ref{eq:bssn_gamma}.  For discrete time
evolution, we use the Iterated Crank-Nicholson method~\cite{Teukolsky00}.
In order to achieve a completely second order method in both space and time
for the coupled system of equations (the Einstein field equations and the
relativistic hydrodynamics equation), we use the time stepping method
described in Figure~\ref{fig:icn_couple}.

\subsection{gauge choices and boundary conditions}

In the 3+1 initial value formulation of general relativity, 
one is free to specify the slicing and spatial coordinate conditions
by specifying the lapse $\alpha$ and shift $\beta^i$, respectively.  
The code described in the previous subsections has been written 
allowing for an arbitrary choice of these gauge variables.  
As described in the next section, each configuration in the CFQE-sequence
approximation has a vanishing trace of the extrinsic curvature
$K$.  As we would like to compare our full general relativistic simulations
to the CFQE-sequence approximation in an invariant manner, it is
desirable to use the same slicing condition during the full 
numerical simulation as that of the CFQE-sequence approximation.
To this end, we would like to select the lapse function $\alpha$ such
that the trace of the extrinsic curvature $K$ remains 0.  Notice
that if one sets $\partial K / \partial t = 0$ in Eq.~\ref{eq:bssn_trk},
then the equation becomes an elliptic equation for the lapse
function $\alpha$.  We have thus implemented a multigrid 
solver~\cite{BrandtA94} for efficiently solving this equation, also known as
the maximal slicing condition~\cite{York79}.  However, solving 
an elliptic equation at every timestep can be numerically expensive. 
We therefore also implement
a variant of the so-called ``1+log'' slicing condition for the lapse,
\begin{equation}
\frac {\partial \alpha}{\partial t} = - 2 \alpha K.
\label{eq:1pluslog}
\end{equation}
Note that this is a completely local condition, and is therefore 
computationally inexpensive.  We use both the maximal slicing condition
and the ``1+log'' slicing condition for simulations presented in
this paper.  For each result, we indicate which 
slicing condition is used.

For the conditions on the shift, we use a slight modification of the
``Gamma-Freezing'' shift equation~\cite{Alcubierre02a}.  Specifically,
we implement the first integral form of the hyperbolic
Gamma-driver (Eq.~46 of reference~\cite{Alcubierre02a}),
\begin{equation}
\frac {\partial {\beta}^i} {\partial t} = C_1 {\tilde{\Gamma}}^i - 
   C_2 {\beta}^i,
\label{eq:shiftcondition}
\end{equation}
where we set the constants $C_1 = C_2 = 0.8$ for all numerical
simulations in this paper.  We find that the Gamma-Freezing 
condition in this modified form, despite its simplicity, results
in enhanced stability
for the type of problems studied in this paper, 
and was also very computationally efficient.

The problem of 
boundary conditions in numerical relativity is only now gaining the
attention it deserves~\cite{Friedrich99,Szilagyi02b,Calabrese01a,Bardeen02,
Szilagyi02a}.  Here, we implement a simple, yet empirically stable, 
boundary condition for the fields describing the gravitational
degrees of freedom, Eqs.~\ref{eq:bssn_phi}-~\ref{eq:bssn_gamma},
due to Alcubierre~\cite{Alcubierre02a}.  Let $f(t,x,y,z)$ represent
a field on which we wish to impose this boundary condition in, say,
the $x$-face of our computational boundary (the $y$- and $z$-face
are similar;  averaging this method yields boundary conditions for
the edge and corner points of the computational boundary).
We take as an ansatz the form for $f$ as $r \rightarrow \infty$
\begin{equation}
f(t,x,y,z) \rightarrow f_{\infty} + \frac{u(r-t)}{r} +
   \frac{w(r+t)}{r}.
\end{equation}
Taking the derivative of this expression with respect to $t$ and $x$, and
eliminating $u$ and $u^\prime$ yields
\begin{equation}
\frac{x}{r} \frac{\partial f}{\partial t} + \frac {\partial f}{\partial x}
+ \frac {x}{r^2} (f-f_\infty) = \frac {1}{r^2} H
\label{eq:miguel_boundary}
\end{equation}
where the function $H = 2 x w^\prime$.  Eq.~\ref{eq:miguel_boundary} is
finite differenced in a second order fashion to obtain a boundary
condition on the field $f$, where we have interpolated the function
$H$ from interior points, assuming a falloff of $\sim \frac {1}{r^2}$ for
$H$.  We adopt this boundary condition for all of the fields describing
the gravitational degrees of freedom.  The boundary condition
used for the hydrodynamical variables is a simple outflow boundary
condition.  Note that all of the hydrodynamical fields are 
trivially small at the computational boundary (the atmosphere has
a baryonic mass density that is $10^9$ times smaller than
the central baryonic mass density of the neutron stars), thus the boundary
conditions used for these fields are relativity unimportant.   

\section{The Conformally Flat, Quasi-Equilibrium (CFQE)
Approximation}
\label{sec:cfqe}

The mathematical assumptions that go into the 
conformally flat, quasiequilibrium approximation to general 
relativity for binary, corotating neutron stars
are as follows (for the physical motivation
behind the assumptions, see, e.g.,~\cite{Baumgarte98b}).

\begin{itemize}
\item  The physical 3-metric $\gamma_{ij}$ is assumed to be conformally flat
\begin{equation}
\gamma_{ij} = {\psi}^4 \delta_{ij}.
\end{equation}
\item The Lie derivative of the conformal metric ${\psi}^{-4} \gamma_{ij}$ with
respect to the time variable $t$ is identically zero:
\begin{equation}
{\cal L}_t ({\psi}^{-4}\gamma_{ij}) = 0.
\label{eq:cfqe_confmet_vanish}
\end{equation}
\item The trace of the extrinsic curvature and it's time derivative vanishes:
\begin{eqnarray}
K & = & 0 \\
{\cal L}_t \left ( K  \right ) & = &0.
\end{eqnarray}
\item There exists an approximate timelike helical Killing 
vector field $\xi^\mu$ 
\begin{equation}
\xi = \left ( \frac {\partial}{\partial t} \right ) + \Omega 
  \left ( \frac {\partial}{\partial \phi} \right )
\label{eq:killing}
\end{equation}
for some constant $\Omega$.
\item The 4-velocity of the fluid $u^\mu$ is proportional to the approximate
Killing vector field $\xi^\mu$:
\begin{equation}
u^\mu \sim \xi^\mu.
\label{eq:ukilling}
\end{equation}
\end{itemize}
From Eq.~\ref{eq:cfqe_confmet_vanish}, the extrinsic curvature $K_{ij}$ takes
the form
\begin{equation}
K_{ij} = \frac {1}{2 \alpha} \left (
   {\cal D}_i \beta_j + {\cal D}_j \beta_i - \frac {2}{3} \gamma_{ij}
   {\cal D}_k \beta^k \right )
\end{equation}
Using this, along with the assumptions of the conformally flat, 
quasiequilibrium approximation above, the Hamiltonian
constraint, Eq.~\ref{eq:ham_constraint}, and the momentum 
constraints, Eqs.~\ref{eq:mom_constraint}, can be written
as
\begin{equation}
{\partial}^i{\partial}_i \psi + \frac {1}{8 \psi^{7}} \tilde{K}_{ij}
\tilde{K}^{ij} + 2 \pi \psi^5 (\rho h W^2 -P) = 0
\label{eq:cfqe_ham}
\end{equation}
and
\begin{equation}
{\partial}^j{\partial}_j \beta^i + 
   \frac {1}{3} {\partial}^i {\partial}_j \beta^j -
   \tilde{K}^{ij} {\partial}_j \left ( \frac {2 \alpha}{\psi^6} \right ) -
   16 \pi \alpha \psi^4 \rho h W^2 v^i = 0,
\label{eq:cfqe_mom}
\end{equation}
respectively,  where we define the conformal extrinsic 
curvature $\tilde{K}_{ij}$ as
\begin{equation}
\tilde{K}_{ij} = \psi^2 K_{ij}.
\end{equation}
Using this form of the Hamiltonian constraint, the maximal 
slicing condition, ${\cal L}_t K = 0$, can be written as
\begin{equation}
\partial^i \partial_i (\alpha \psi) - 
   \frac {7 \alpha}{8 \psi^8} \tilde{K}_{ij} \tilde{K}^{ij} +
   2 \pi \psi^4 ( 2 \rho (1 + \epsilon) - 3 P - 3 \rho h W^2) = 0.
\label{eq:cfqe_maximal}
\end{equation}
Since the quasiequilibrium approximation assumes the existence of a 
timelike Killing vector, we can analytically find the first integral of
the relativistic Bernoulli equation, which is simply
\begin{equation}
\frac {u^t}{h} = \verb+const+.
\end{equation}
Along with the normalization condition for the 4-velocity, 
$u^\mu u_\mu = -1$, this equation can be explicitly written as
\begin{equation}
h^2 \left ( \alpha^2 - \psi^4
   ( {(\beta^x - y \Omega)}^2 + {(\beta^y + x \Omega)}^2 + {(\beta^z)}^2 
   \right ) = \verb+const+.
\label{eq:bernoulli}
\end{equation}

\subsection{CFQE configurations}

\begin{figure}
\vspace{0.0cm}
\hspace{0.0cm}
\psfig{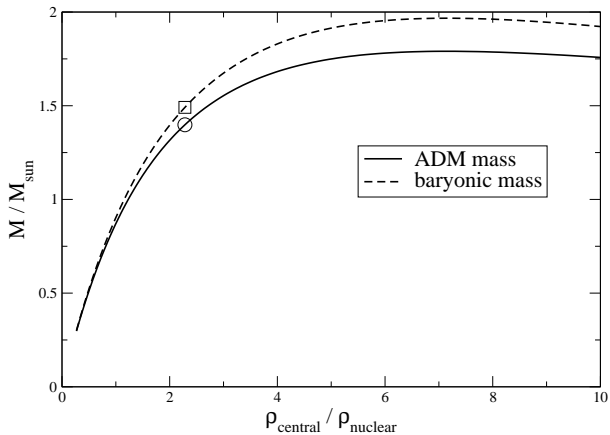}
\caption{
The ADM mass and baryonic mass (in units of $M_\odot$) as 
a function of central density (in units of nuclear density)
for single, static neutron star configurations.
All studies in this paper are done with stars of baryonic mass
$1.49 M_\odot$ (square).  The corresponding ADM mass of this
single static configuration is $1.4 M_\odot$ (circle).
}
\vspace{0.0cm}
\label{fig:tov}
\end{figure}

We use the algorithm detailed in~\cite{Baumgarte98b}, along with 
a parallel multigrid solver,  to
simultaneously solve the algebraic
Bernoulli integral equation, Eq.~\ref{eq:bernoulli}, and the 
five elliptic equations corresponding to the Hamiltonian
constraint (Eq.~\ref{eq:cfqe_ham}), the three momentum 
constraints (Eqs.~\ref{eq:cfqe_mom}), and the maximal slicing
equation (Eq.~\ref{eq:cfqe_maximal}), for the conformal factor $\psi$,
the three components of the shift vector $\beta^i$, the 
lapse $\alpha$, and the matter fields, thus producing a single
CFQE configuration.  In numerically solving these elliptic 
equations, we use identical boundary conditions to 
those in~\cite{Baumgarte98b}. We have also assumed a polytropic 
equation of state,
\begin{equation}
P = (\Gamma - 1) \rho \epsilon = k \rho^\Gamma,
\end{equation}
where $k$ is the polytropic constant and $\Gamma$ is the adiabatic index.
The baryonic mass of
each neutron star in a CFQE configuration, $M_0$, is defined
as half of the total rest mass of the configuration,
\begin{equation}
M_0 = \frac {1}{2} \int d^3\!x \, \sqrt{\gamma} \rho W.
\end{equation}
In this paper, we set $\Gamma = 2$ and $k = 0.0445 \frac {c^2}{\rho_n}$,
where $\rho_n$ is nuclear density 
(approximately $2.3$  x  $10^{14} \; g/cm^3$). For these values of parameters,
a single static neutron star configuration that is stable has
a maximum 
ADM mass of $1.79 M_{\odot}$ and a baryonic mass
of $1.97 M_{\odot}$ ($M_\odot$ is 1 solar mass).  
For the studies in
this paper, we use neutron stars with a baryonic mass of $1.49 M_\odot$
each, which
is approximately $75\%$ that 
of the maximum stable configuration.  The ADM mass of a single static 
neutron star for this configuration 
is $1.4 M_\odot$ (see Figure~\ref{fig:tov}).

Once the baryonic mass of each of the stars is fixed,
the only remaining degree of freedom in the specification of a
CFQE corotating configuration is the
separation of the two stars.  A
natural invariant way of specifying the separation of the two 
neutron stars is to calculate the geodesic distance on the constant $t$ hypersurface between the points
in each of the neutron stars that corresponds to the maximum 
baryonic mass density.  Specifically, if $x^i_{M1}$ and $x^i_{M2}$ 
are the spatial coordinates of the points of maximum baryonic mass density in
the first and second star, respectively, then the geodesic whose length
will represent the (spatially) invariant separation of the neutron stars is
the curve
\begin{equation}
X^i(\lambda), \:\:\: \lambda \in [0,1],
\end{equation}
where
\begin{equation}
X^i(\lambda\!=\!0) = x^i_{M1}, \:\:\: X^i(\lambda\!=\!1) = x^i_{M2},
\end{equation}
such that
\begin{equation}
\frac {d^2 X^i}{d \lambda^2} + {\Gamma^i}_{jk} 
   \frac {dX^j}{d \lambda} \frac {dX^k}{d \lambda} = 0.
\label{eq:geodesic}
\end{equation}
The geodesic distance $\ell_{1,2}$ between the two stars is now defined as
\begin{equation}
\ell_{1,2} = \int_{\lambda=0}^{\lambda=1} d\lambda \: 
   \sqrt{ \gamma_{ij} \frac {dX^i}{d \lambda} \frac {dX^j}{d \lambda} }
\label{eq:geolength}
\end{equation}
We use a relaxation technique for numerically 
solving the geodesic ODE, Eq.~\ref{eq:geodesic}, 
and then use a standard quadrature formula for evaluating the 
integral for $\ell_{1,2}$, Eq.~\ref{eq:geolength}.  We 
choose the discretization for solving Eq.~\ref{eq:geodesic} and
Eq.~\ref{eq:geolength} to be 10 times as fine as the 3D discretization, 
and use a third order accurate 
interpolator to calculate the Christoffel symbols
${\Gamma^i}_{jk}$ from the 3D grid.  The maximum baryonic mass points,
$x^i_{M1}$ and $x^i_{M2}$, are located by finding the maximum of a 3D
third order accurate interpolation polynomial, the interpolation
being centered on the discrete point on the 3D grid that has the
largest value of the baryonic mass density in each star.

\begin{figure}
\vspace{0.0cm}
\hspace{0.0cm}
\psfig{figure=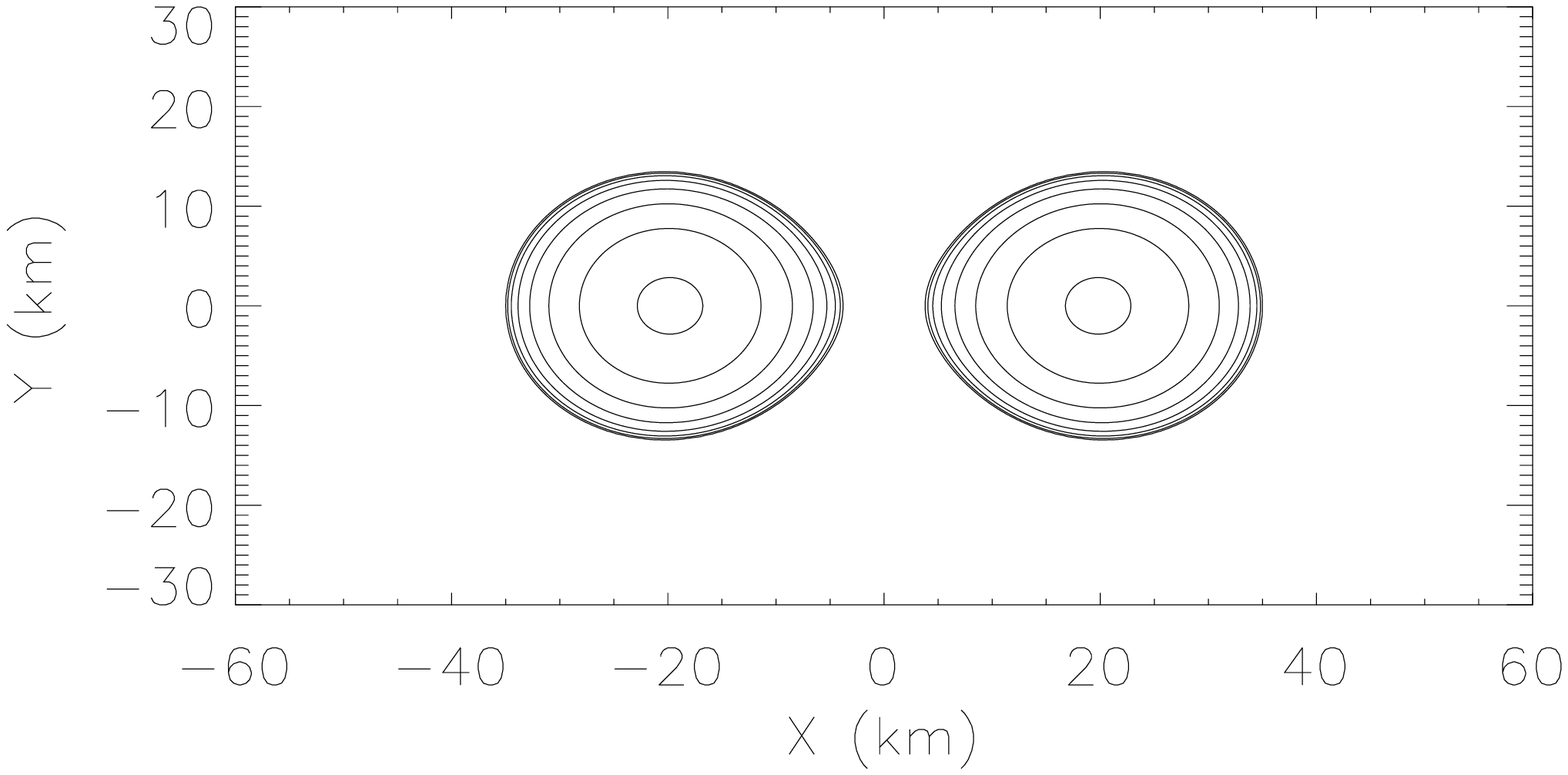,width=8cm}
\psfig{figure=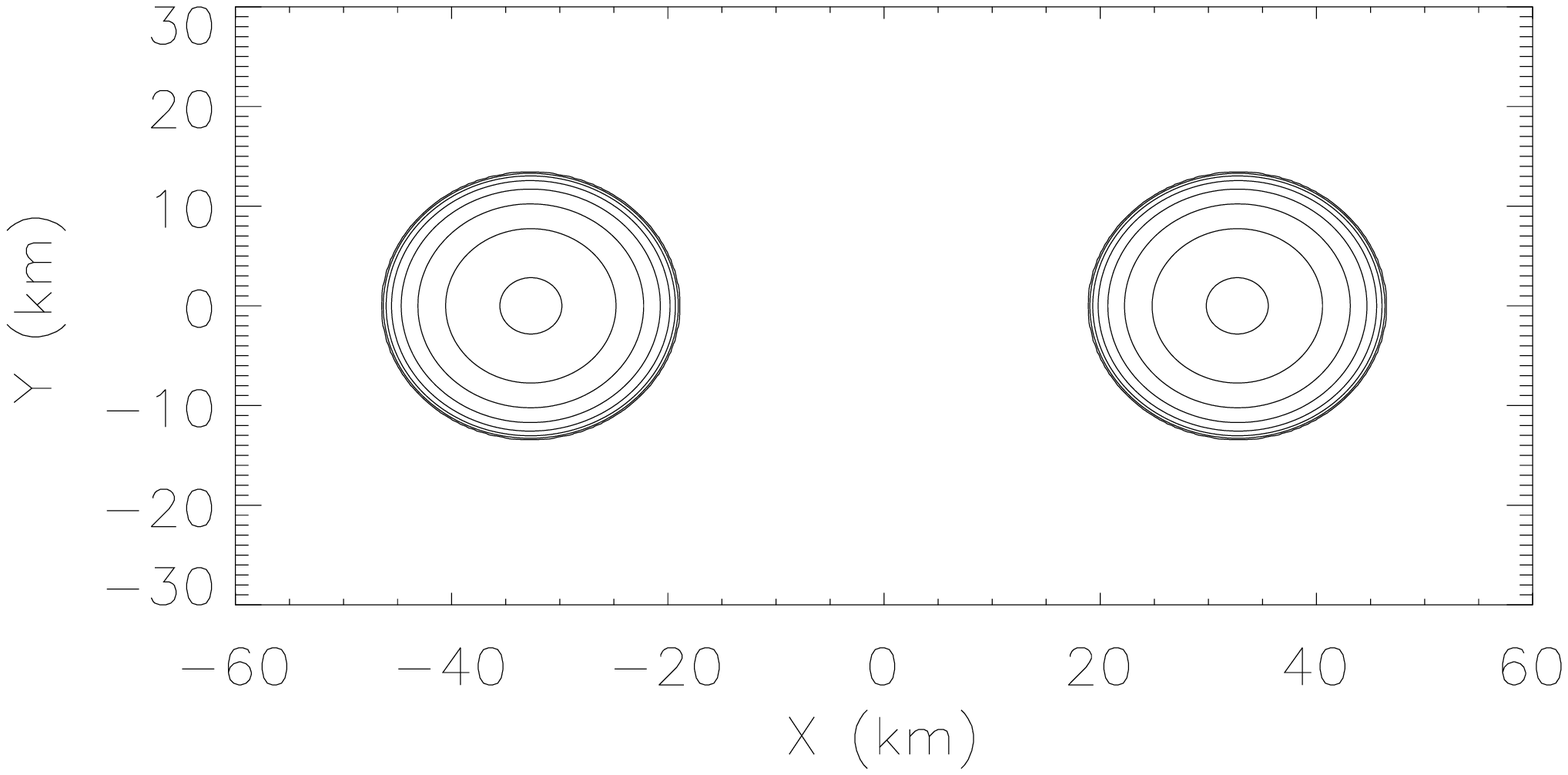,width=8cm}
\caption{Contours of the baryonic mass density for two CFQE configurations,
taken in the equatorial ($z=0$) plane.  The ADM mass of each neutron
star in isolation is $1.4 M_\odot$.
The value of the mass density for the first contour 
is $0.9$ that of the maximum mass
density of the star, with the others decreasing by a factor of 
$0.5$ each.  The CFQE configuration in the top panel has a
geodesic separation of $\ell_{1,2} / M_0 \: = \: 23.44$ (which 
is $51.70$ km), and the configuration in the bottom panel has
a geodesic separation of $\ell_{1,2} / M_0 \: = \: 35.72$ (which 
is $79.00$ km).
These configurations represent the smallest and largest separation
CFQE configurations that are used as initial data for the fully consistent
general relativistic numerical calculations done in this paper.
}
\vspace{0.0cm}
\label{fig:cfqe_restmass}
\end{figure}

In Figure~\ref{fig:cfqe_restmass}, we plot logarithmically
spaced contours of the baryonic mass density in the 
equatorial ($z=0$) plane for two 
CFQE configurations
representing the smallest and largest separations used as 
initial data for the 3D numerical simulations performed in 
this paper.  

\subsection{The CFQE-sequence approximation}
\label{sec:cfqesequence}
A CFQE-sequence is constructed by stringing together several constant
baryonic mass CFQE configurations, each configuration differing in only
the separation of the neutron stars.  The idea is that, due to the
fact that the time scale of the gravitational radiation, and thus, 
that of the orbital decay, is much longer than the orbital timescales,
the binary neutron stars are considered to be in ``quasiequilibrium''.
Thus, at any particular time, a binary neutron 
star system can be described by one particular CFQE configuration;
the effect 
of the gravitational 
radiation is, over the timescales of one orbit, to alter the
configuration to a new CFQE configuration with a slightly smaller
separation.

It is typical in CFQE approximation studies to calculate an effective
binding energy for each configuration.  Following~\cite{Baumgarte98b},
we define the (dimensionless) effective 
binding energy $E_b$ of a single configuration to
be
\begin{equation}
E_b = \frac {M_{ADM} - 2 M_{NS\infty}} {M_0}
\label{eq:bindingenergy}
\end{equation}
where $M_{ADM}$ is the ADM mass of the configuration, and $M_{NS\infty}$
is the ADM mass of a single static neutron star in isolation with rest
mass $M_0$.  The usual expression for the ADM mass of an asymptotically
flat spatial slice is
\begin{equation}
M_{ADM} = \frac {1}{16 \pi} \lim_{r \rightarrow \infty} 
   \sum_{i,j=1}^{3}  \oint dA \, 
   \left ( \frac {\partial \gamma_{ij}}{\partial x^i} -
           \frac {\partial \gamma_{ii}}{\partial x^j} \right ) N^j,
\end{equation}
where $N^i$ is the unit outward normal to the sphere of constant 
radius $r$, which is the domain of integration.  Using the fact that
the CFQE configurations are conformally flat,
the ADM mass of any configuration reduces to the volume integral
\begin{equation}
M_{ADM} = - \frac {1}{2 \pi} \int d^3\!x \: \left (
   \partial^i \partial_i \psi \right ).
\end{equation}

\begin{figure}
\vspace{0.0cm}
\hspace{0.0cm}
\psfig{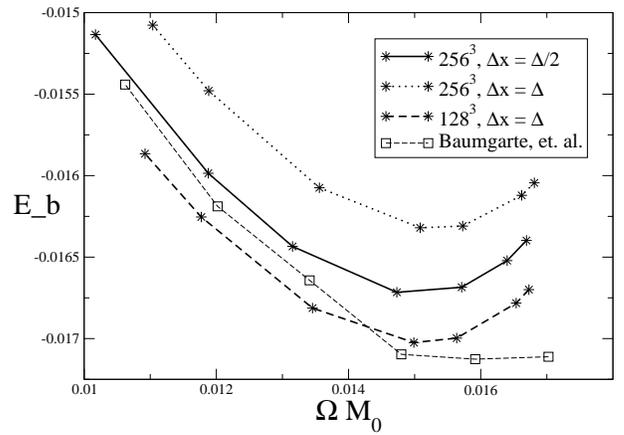}
\caption{
The effective binding energy $E_b$ as a function of the orbital
angular velocity parameter $\Omega$ for 
constant baryonic mass CFQE-sequences.  The baryonic mass of each of the
neutron stars is $1.49 M_\odot$.  Shown are results for our CFQE
configuration solver at different resolutions and outer boundary placements.
Tables IV, V, and VI of reference~\cite{Baumgarte98b} were interpolated
to obtain the results from Baumgarte, et. al.  
}
\vspace{0.0cm}
\label{fig:cfqe_seq_data}
\end{figure}

In Figure~\ref{fig:cfqe_seq_data}, we plot the effective binding 
energy $E_b$ as a function of the angular velocity parameter $\Omega$,
for a constant baryonic mass CFQE-sequence (each star has a baryonic 
mass of $1.49 M_\odot$).  The highest value of $\Omega$ for each
of our calculated CFQE-sequences corresponds to a CFQE configuration where
the neutron stars are touching.  The configuration corresponding
to the minimum effective binding energy $E_b$ is the ISCO configuration.
It is this point along the CFQE-sequence approximation where the 
quasiequilibrium configuration goes secularly unstable;  the
subsequent evolution of the system is thought to enter a ``plunge phase''
where the neutron stars coalesce within an orbital timescale.

\begin{figure}
\vspace{0.0cm}
\hspace{0.0cm}
\psfig{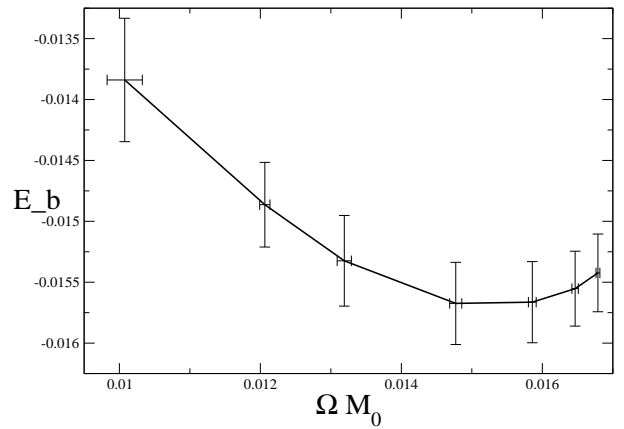}
\caption{
The effective binding energy $E_b$ as a function of the orbital
angular velocity parameter $\Omega$.  The data in this
plot were obtained by using Richardson extrapolation
from the data in Figure~\ref{fig:cfqe_seq_data}.  The form of 
the truncation and boundary error are given by Eq.~\ref{eq:cfqe_errfunc}.
The error bars were computed using Eq.~\ref{eq:cfqe_errorbar}.  These
results represent what one would obtain in the limit as both
$\Delta x \rightarrow 0$ and the location of the 
outer boundary goes to $\infty$.
}
\vspace{0.0cm}
\label{fig:cfqe_richard}
\end{figure}

Of course, when using a computer code to solve 
differential equations, results 
produced at one single
resolution are meaningless due to the fact that we have no way
of knowing how big the numerical errors are.
It is always important to make numerical calculations at different 
resolutions, in order to be able to assess the numerical 
errors that are inherent to any discretization.  Not only do we have
the usual truncation errors (which are due to the fact that higher
order terms in the Taylor expansion of functions have been dropped
in our finite difference approximation)
of any finite difference approximation, we also have boundary errors,
as we are solving a set of 5 coupled elliptic equations to arrive at
any CFQE configuration.  Therefore, we must make at least three 
numerical experiments in order to assess the effect of both resolution
and boundary placement on the numerical results.  We show our 
CFQE configuration results
for three combinations of resolution and outer boundary placement,
alongside the 
results obtained by Baumgarte,~et.~al.~\cite{Baumgarte98b}.
Our CFQE sequence results are in fairly good agreement with
those of~\cite{Baumgarte98b}, considering the two terms that
are subtracted in
the construction of $E_b$, Eq.~\ref{eq:bindingenergy},
are the same to the first two or three significant digits.
The resolution used in the two studies are comparable.   However,
in~\cite{Baumgarte98b}, the number of gridpoints across the neutron star
was kept fixed.  We, on
the other hand, keep (for each sequence calculated) the boundary fixed, and
thus have different numbers of gridpoints across the star in each
sequence.  Notice that, as we both increase the resolution and
increase the distance from the center of mass to the boundary of 
the computational domain, the shape of the curves in 
Figure~\ref{fig:cfqe_seq_data} remain roughly the same.  
In fact, we can perform a Richardson extrapolation on our numerical
results by fitting each point on the curve to an error function.
For example, we can use the error function
\begin{equation}
(E_b)_{numerical} = (E_b)_{exact} + C_1 {(\Delta x)}^2 + 
   \frac {C_2}{{r_{id}}^2},
\label{eq:cfqe_errfunc}
\end{equation}
where $(E_b)_{numerical}$ is the numerical value for the binding
energy $E_b$ obtained using a 
specific discretization $\Delta x$ and coordinate distance
from the center of mass to the outer boundary $r_{id}$ 
(we use $r_{id}$ to denote this distance in solving the
elliptic equations for the CFQE configurations, and we will use
$r_b$ to denote the coordinate distance from the center of mass to the outer
boundary of the computational domain used during the dynamical 
evolutions.  Of course, we always have $r_{id} \geq r_b$.).
Note that the first nonzero term for the expansion of the boundary
error does not contain a $1/r_{id}$ term.  This is due to that fact
that the boundary conditions we use in solving the elliptic equations
for the CFQE configurations, which are identical to those 
used in~\cite{Baumgarte98b}, are exact to this order.
$(E_b)_{exact}$ is
the binding energy in the limit as $\Delta x \rightarrow 0$ and 
$r_{id} \rightarrow \infty$, i.e. that given by an exact solution to the 
differential equations.  Of course, it is $(E_b)_{exact}$ that we
are interested in.  We use our three CFQE-sequences shown in 
Figure~\ref{fig:cfqe_seq_data} to solve for 
the three unknowns $(E_b)_{exact}$, $C_1$,
and $C_2$ from Eq.~\ref{eq:cfqe_errfunc}.  A generous estimate of the
total error, which will be used for the size of the error bars, is
\begin{eqnarray}
\verb+error+ = \verb+max+\{ &
   \left | C_1 {(\Delta x)^2}_{best} \right |,
   \left | \frac {C_2}{{{r_{id}}^2}_{best}} \right |, \nonumber \\
   & \left | C_1 {{(\Delta x)}^2}_{best} + 
   \frac {C_2}{{{r_{id}}^2}_{best}} \right | 
   \}.
\label{eq:cfqe_errorbar}
\end{eqnarray}
The results are plotted in Figure~\ref{fig:cfqe_richard}. 

\subsection{neutron star spin corrections to the CFQE-sequence approximation}
\label{sec:nsspin}

Note in Figure~\ref{fig:cfqe_richard} that the effective binding
energy attains a minimum as the separation between the neutron
stars decreases.   As explained in the introduction, 
the CFQE-sequence approximation
thus predicts a secular instability at this orbital separation,
at which point
the evolution of the system will change from
quasistationary into a plunge phase, where the neutron
stars would merge on timescales of the orbital period.
Several types of arguments have been used to support this
claim~\cite{Baumgarte98b,Baumgarte98c}.  On the one hand, it is
intuitively clear that, in the presence of a dissipation mechanism
(gravitational waves are slowly dissipating the binding energy
of the binary system), if the binding energy would increase with
decreasing separation, this would energetically be an 
unstable situation.  There are also turning point 
theorems~\cite{Sorkin82,Friedman88,Baumgarte98c} 
in full nonlinear general relativity which state that
if a sequence of equilibrium configurations attain a
minimum in the ADM mass-energy, then this equilibrium configuration
is secularly unstable.  Hence, to the accuracy of the 
CFQE-sequence approximation,
the minimum in Figure~\ref{fig:cfqe_richard} is 
often referred to as the innermost stable circular
orbit (ISCO) configuration.

One problem with the view that these CFQE-sequences somehow approximate
fully consistent solutions to the Einstein equations coupled to 
the relativistic hydrodynamic equations has to do with the spin of the
the individual neutron stars.  Note that in each 
of the CFQE configurations,
the 4-velocity of the fluid is assumed proportional to the
timelike helical Killing vector field.  The neutron stars are thus spinning at
the exact same frequency as the orbital frequency, $\Omega$.  The 
neutron stars are said to be corotating with the orbital motion.  
However, it has been known for some time that
realistic binary neutron stars cannot be tidally locked during
the late stages of inspiral (i.e., less than 1000 orbits 
until the final plunge)~\cite{Bildsten92}.  

This raises a question:  if a CFQE configuration is used as 
initial data,
should we expect the subsequent solution to the Einstein 
equations to follow the CFQE-sequence approximation,
and keep the stars tidally locked as the stars slowly inspiral?  In light
of the results of~\cite{Bildsten92}, 
the answer must be no.  To first order, we
would expect the stars to retain approximately the same spin during 
the dynamical evolutions as is given in the CFQE configuration that
was used as initial data.  

\begin{figure}
\vspace{0.0cm}
\hspace{0.0cm}
\psfig{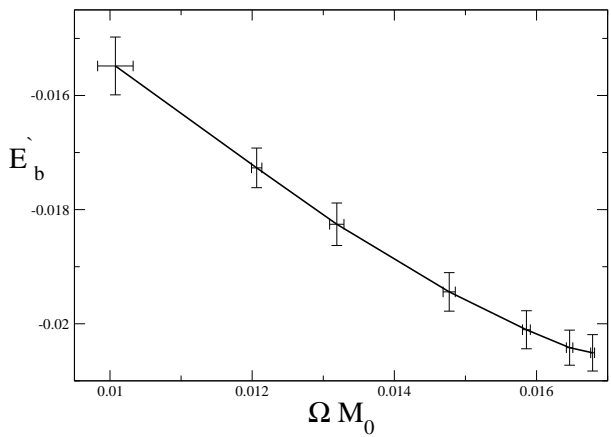}
\caption{
The effective binding energy $E_b^\prime$,
Eq.~\ref{eq:newbindingenergy}, as a function of the orbital
angular velocity parameter $\Omega$.  The method for
obtaining the data and error bars in this plot
are equivalent to those used in producing
Figure~\ref{fig:cfqe_richard}.
}
\vspace{0.0cm}
\label{fig:cfqe_richard_new}
\end{figure}

The resolution of this first question presents a second:  what is the
status of the ISCO?  If the stars do {\it not}, in fact, stay tidally
locked during dynamical evolution, will this affect the location and/or
existence of a turning point in the binding energy curve, i.e. the
CFQE ISCO configuration?  One can gain insight into this question
by comparing the spin kinetic energy of relativistic, uniformly rotating
neutron stars to the energy scales involved in calculating the
binding energy, Eq.~\ref{eq:bindingenergy}.  To this end, we define
a new effective binding energy that explicitly takes the spin
kinetic energy of the neutron stars into account:
\begin{equation}
E_b^\prime = \frac {M_{ADM} - 2 M_{NS\infty} - 
2 \, \Delta \! M_{RNS\infty}(\Omega)} {M_0}
\label{eq:newbindingenergy}
\end{equation}
where $\Delta\!M_{RNS\infty}(\Omega)$ is the difference between 
the ADM mass of an isolated stationary neutron star with baryonic 
mass $M_0$ uniformly rotating with 
angular velocity $\Omega$ and the ADM mass of an isolated static non-rotating
neutron star with baryonic mass $M_0$.   Thus, $\Delta\!M_{RNS\infty}(\Omega)$
represents the spin kinetic energy of a uniformly rotating neutron
star rotating with an angular velocity $\Omega$.
We plot this new effective binding
energy $E_b^\prime$ in Figure~\ref{fig:cfqe_richard_new}.  As can be 
seen, the binding energy where the spin kinetic energy of each of the neutron
stars has been manually factored out no longer attains a minimum
(the data point with the highest value of $\Omega$ corresponds to
the CFQE configuration in which the two stars are touching).  This
indicates that the secular stability of very close binary neutron
stars could depend very sensitively on the spin-orbit coupling of
the binary system.  The two extreme cases, that of tidally
locked neutron stars and non-rotating neutron stars, are depicted
by Figures~\ref{fig:cfqe_richard}~and~\ref{fig:cfqe_richard_new},
respectively.  The fact that we expect a weak spin-orbit coupling
suggests that, in the fully consistent general relativistic 
hydrodynamics simulations, where we do not expect the neutron
stars to remain tidally locked,  there may be no turning point and
that there may not, in fact, be any ISCO configuration.  

\section{Comparing Numerical Evolution in Full General Relativistic Theory
with the CFQE-sequence approximation}
\label{sec:short_evolve}

Using our numerical evolution code described in Section~\ref{sec:formulation}, 
along with the CFQE configurations described in Section~\ref{sec:cfqe} as
initial data, we analyze to what degree the CFQE configurations actually
correspond to quasiequilibrium configurations in full general
relativity.  We do this by computing fully
consistent general relativistic evolutions numerically using CFQE
configurations as initial data, and comparing the resulting solution 
of the Einstein field equations with the spacetime obtained by the CFQE
approximation.  Recall that one major assumption in the CFQE approximation
method is that the change of the binary separation is small on
timescales less than one orbital period.  If the solution to the
Einstein equations obtained by using any particular CFQE configuration
as initial data deviates significantly from the CFQE-sequence spacetime 
on suborbital timescales, then that particular CFQE configuration
cannot be thought of as representing a quasiequilibrium configuration.
Of course, the quasiequilibrium approximation becomes better as the 
separation of the neutron stars increases.  That question then
becomes:  how close can the neutron stars be before the CFQE approximation
breaks down?  In this section, we describe a generic method that
we subsequently use to answer this question.  In short, we use
CFQE configurations of differing initial separation as initial 
data for our fully consistent general relativistic code, and 
compare the resulting solutions to the Einstein field equations
to the CFQE-sequence spacetime in a coordinate independent fashion.
The magnitude of these differences, which we find to decrease as
the initial separation of the neutron stars increases (as expected), will
therefore quantify the degree of failure of the CFQE configurations to be 
truly quasiequilibrium configurations in full general relativity.

Recall that the
CFQE-sequence approximation is done in a maximally sliced $(K = 0)$
coordinate system. In order to facilitate a meaningful
comparison between the fully general relativistic calculation
and the CFQE-sequence approach, we adopt this slicing condition
for all numerical computations done in this section.
Thus, we determine the lapse function $\alpha$ by solving
the maximal slicing condition, which is the elliptic equation
obtained by setting $\partial K / \partial t$ to $0$ in
Eq.~\ref{eq:bssn_trk}.
In order to decrease the computational cost involved
in solving this elliptic equation, we only solve the maximal
slicing condition every eight timesteps.  We use 
``1 + log'' evolution, Eq.~\ref{eq:1pluslog}, for the lapse $\alpha$
for the timesteps where we do not solve the maximal slicing 
condition.  We have performed several numerical tests
where we do, in fact, solve the maximal slicing condition for
every timestep, and have seen only a negligible affect on
the results (in fact, the measures of the violation of the 
Einstein equations by the CFQE approximation that we find in 
this section
slightly {\it increase} 
when we solve the maximal slicing condition for the lapse $\alpha$ 
at every timestep).

In order to meaningfully compare our fully consistent 
general relativistic simulations with the CFQE-sequence approximation,
we must compare quantities that are independent of our choice of
coordinates.  Since we are using the same slicing condition as that 
assumed by the CFQE-sequence approximation (namely, maximal slicing),
we must compare our numerical results with the CFQE-sequence
approximation using quantities that are either fully 4-invariant quantities,
or 3-invariant quantities (quantities that are invariant under
spatial coordinate transformations).  In the following, we will 
construct measures for comparing our numerically constructed
spacetime with the CFQE-sequence spacetime
based on both the conformal flatness assumption and
the assumption of the existence of a timelike helical Killing vector field.
Both of these measures will be exactly zero for the CFQE-sequence spacetime,
but will not necessarily be zero for the full solution to the
Einstein field equations using CFQE configurations as initial data. 
The magnitude of these measures will thus quantify the magnitude
of the CFQE configuration's failure to correspond to a true 
quasiequilibrium configuration in full general relativity.
In Table~\ref{tab:shortconfig}, we show the parameters for the 
CFQE configurations that are used as initial data for the 
simulations done in this section.  

\begin{table}
\begin{tabular}{|c|c|c|c|c|c|}  \hline \hline
\hspace{0.0cm} { Config.} \hspace{0.0cm}  &
\hspace{0.0cm} {$M_0 / M_\odot$} \hspace{0.0cm}  &
\hspace{0.0cm} {$\ell_{1,2} / M_0$} \hspace{0.0cm}  &
\hspace{0.0cm} {$J / {M_0}^2$} \hspace{0.0cm}  &
\hspace{0.0cm} {$M_{ADM} / M_0$} \hspace{0.0cm}  &
\hspace{0.0cm} {$\Omega M_0$} \hspace{0.0cm} \\ \hline \hline
{NS-1} &
   {1.490} &
   {23.44} &
   {3.770} &
   {1.857} &
   {0.01547} \\ \hline
{NS-2} &
   {1.490} &
   {25.94} &
   {3.762} &
   {1.857} &
   {0.01296} \\ \hline
{NS-3} &
   {1.490} &
   {29.78} &
   {3.812} &
   {1.858} &
   {0.01022} \\ \hline
{NS-4} &
   {1.490} &
   {35.72} &
   {3.934} &
   {1.859} &
   {0.007460} \\ \hline \hline
\end{tabular}
\vspace{0.0mm}
\caption{CFQE configuration parameters used as initial data 
for fully consistent general relativistic simulations in 
Section~\ref{sec:short_evolve}.  For reference, the ISCO configuration
has a proper geodesic separation of $\ell_{1,2} = 24.0 \: M_0$.
}
\label{tab:shortconfig}
\end{table}

\subsection{The Killing vector field assumption of the CFQE-sequence
approximation}
\label{sec:killing}

Recall from Section~\ref{sec:cfqe} that one assumption of
the CFQE-sequence approximation is the existence of an approximate 
timelike helical Killing
vector field, Eq.~\ref{eq:killing}.  As we are assuming a
corotating matter field, the 
4-velocity of the
fluid must be proportional to this Killing vector field,
Eq.~\ref{eq:ukilling}.
Here, we monitor this quasi-equilibrium (QE) assumption in our fully consistent general
relativistic numerical calculations.

That the 4-velocity $u^a$ be
proportional to a Killing vector field is equivalent to the 
vanishing of a symmetric, type-(0,2) 4-tensor $Q_{ab}$
\begin{equation}
\label{eq:q}
Q_{ab} \equiv \nabla_a u_b + \nabla_b u_a + u_a a_b + u_b a_a
\end{equation}
where $a^a \equiv u^b \nabla_b u^a$ is the 4-acceleration of the
fluid ($\nabla_a$ denotes the covariant derivative operator 
compatible with the 4-metric $g_{ab}$).
Notice that the quantity $Q_{ab} u^a$ vanishes identically.
We can thus monitor the space-space components of $Q_{ab}$ during our 
simulations as a way of monitoring how
well the 4-velocity $u^a$ stays proportional to a 
Killing vector field.  Define $Q_{ij}$ as the projection
of $Q_{ab}$ onto the constant $t$ spatial slice:
\begin{equation}
Q_{ij} = {P_i}^a {P_j}^b Q_{ab} = 
   Q_{ab} {\left ( \frac{\partial}{\partial x^i} \right )}^a
          {\left ( \frac{\partial}{\partial x^j} \right )}^b
\end{equation}
where $P_{ab} = g_{ab} + n_a n_b$ is the projection operator
onto the constant $t$ spatial slices.   The unit normal to 
these spatial slices is $n^a = \alpha t^a + \beta^a$.  In our
Cartesian coordinates $x^i$ for the spatial slices, the
components of $Q_{ij}$ form a 
3x3 matrix.  The norm of this matrix, which itself
is a coordinate-independent quantity, is the square root of the 
largest eigenvalue of $Q_{ij} {Q^j}_k$, where we have raised
and lowered 3-indices with the 3-metric.  In our case, $Q_{ij}$ is
symmetric, and the matrix norm reduces to the largest eigenvalue of
$Q_{ij}$ itself.  We denote this coordinate-independent value
of the norm of $Q_{ij}$ as 
$\left | Q_{ij} \right | $.  Note that if the 4-velocity
is proportional to an exact Killing vector, $\left | Q_{ij} \right | $
will be exactly zero.  Of course, in the fully consistent general 
relativistic treatment, $\left | Q_{ij} \right | $ will not
vanish.  What is required is a sense of the relative size of
$\left | Q_{ij} \right | $.  Notice that in Eq.~\ref{eq:q},
$Q_{ab}$ is constructed out of only two separate (symmetric) parts,
the $\vec{\nabla}\vec{u}$ part and the $\vec{u}\vec{a}$ 
part.  We can therefore naturally normalize $\left | Q_{ij} \right | $
by the norms of these two principle parts.  If we define
\begin{eqnarray}
{Q_1}_{ab} & \equiv & \nabla_a u_b + \nabla_b u_a \\
{Q_2}_{ab} & \equiv & u_a a_b + u_b a_a,
\end{eqnarray}
then a naturally normalized scalar field $Q$ which denotes the deviation from 
the 4-velocity $u^a$ being proportional to a Killing vector 
field is
\begin{equation}
Q = \frac {\left | Q_{ij} \right |} 
          {\mbox{max} \{ \left | {Q_1}_{ij} \right | ,
                         \left | {Q_2}_{ij} \right | \} }.
\end{equation}
This normalization provides a measure for the deviation of  
the 4-velocity $u^a$ from being proportional to a Killing 
vector field (the QE assumption of the CFQE approach);
a value of $Q=0$ signifies that the 4-velocity of the fluid
is exactly proportional to a timelike Killing vector, while 
a value of $Q$ of order unity would signify a significant
violation of the QE assumption.
The monitoring
of $Q$ during a fully consistent general relativistic simulation
is then a quantitative measure of the accuracy of the
QE approximation. 
Since $Q$ is 
meaningful only inside the fluid bodies, a natural global measure 
of the magnitude of $Q$ is its the baryonic mass 
weighted integral, denoted by
$\langle Q \rangle$:
\begin{equation}
\label{eq:weight_q}
\langle Q \rangle = 
   \frac {\int d^3\!x \: \left | Q \right | \sqrt{\gamma} \rho W }
   {\int d^3\!x \: \sqrt{\gamma} \rho W},
\end{equation}
where the integrals are taken to be over the entire spatial slice.

\begin{figure}
\vspace{0.0cm}
\hspace{0.0cm}
\psfig{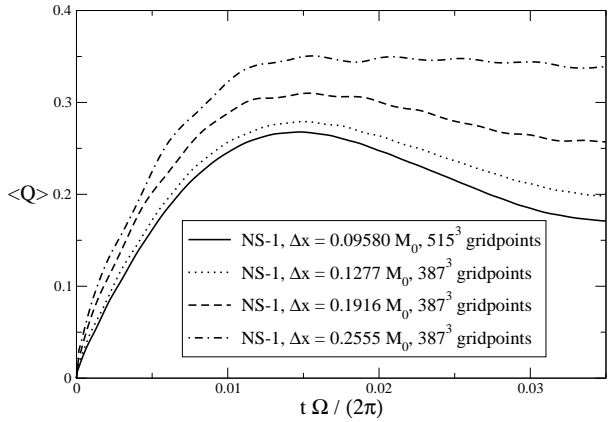}
\caption{
We plot the quantity $\langle Q \rangle$ (Eq.~\ref{eq:weight_q}) as a function
of time
for fully consistent general relativistic numerical
calculations, using CFQE configuration NS-1 as initial data
(see Table~\ref{tab:shortconfig}).  A variety of discretization
parameters $\Delta x$ are used. 
}
\vspace{0.0cm}
\label{fig:q_ns1}
\end{figure}

In Figure~\ref{fig:q_ns1}, we plot $\langle Q \rangle$ as 
a function of time for 
a small fraction of an orbit in the fully consistent general
relativistic numerical simulations.  Configuration NS-1 
was used as initial data (see Table~\ref{tab:shortconfig}).
Various resolutions were used, along with different numbers of gridpoints
for the computational domain.  As stated in the section where
we numerically solved for the CFQE configurations, 
Section~\ref{sec:cfqesequence}, it is important to run any 
simulation at multiple resolutions and boundary placements, in
order to assess the magnitude of the boundary error and the
finite difference truncation error on the numerical results.  

Notice in Figure~\ref{fig:q_ns1} that the value of
$\langle Q \rangle$ appears to be converging to a curve that attains a maximum 
value of approximately $\langle Q \rangle = 0.26$ after  
$1.5\%$ of an orbit ($2\pi/\Omega$ is approximately $1$ orbital 
period).  This is quite a fast growth, and a (perhaps unexpectedly)
large value attained in such a short time.  Recall
that $\langle Q \rangle$ is normalized such that a 
value of $\langle Q \rangle$ of order unity 
represents a significant failure of the 4-velocity $u^a$ to be
proportional to a Killing vector.  
One reason for this rapid 
growth has been discussed in Section~\ref{sec:nsspin}:  there is no
mechanism in the actual evolution 
to spin up the neutron stars in order that they remain in corotation with
the orbital motion.  This corotation condition is ``force-fed'' into
the CFQE configurations.
Of course, one expects that the 
CFQE-sequence approximation should become better
as the separation between the two neutron stars increases.
The orbital angular velocity, and thus, the
spin of the neutron stars, decreases with increasing separation.  
Hence, both the gravitational radiation reaction and the error
introduced by the artificial spin-up of the individual stars
due to the corotation assumption 
are lessened.  These two factors will increase the validity of
CFQE approximation for increased neutron star separations.

\begin{figure}
\vspace{0.0cm}
\hspace{0.0cm}
\psfig{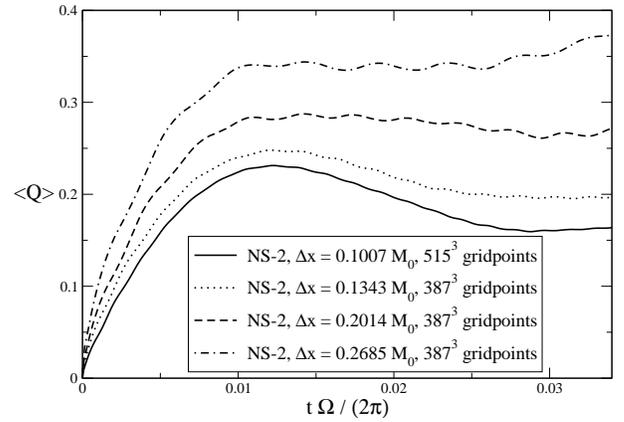}
\caption{
We plot the quantity $\langle Q \rangle$ (Eq.~\ref{eq:weight_q}) as a function
of time
for fully consistent general relativistic numerical
calculations, using CFQE configuration NS-2 as initial data
(see Table~\ref{tab:shortconfig}).  A variety of discretization
parameters $\Delta x$ are used.
}
\vspace{0.5cm}
\label{fig:q_ns2}
\end{figure}

\begin{figure}
\vspace{0.0cm}
\hspace{0.0cm}
\psfig{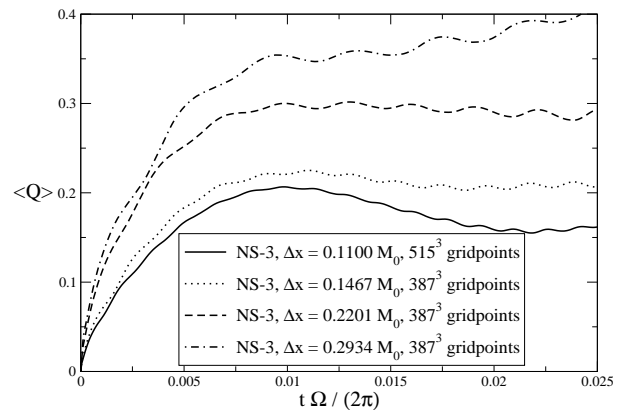}
\caption{
We plot the quantity $\langle Q \rangle$ (Eq.~\ref{eq:weight_q}) as a function
of time
for fully consistent general relativistic numerical
calculations, using CFQE configuration NS-3 as initial data
(see Table~\ref{tab:shortconfig}).  A variety of discretization
parameters $\Delta x$ are used.
}
\vspace{0.5cm}
\label{fig:q_ns3}
\end{figure}

\begin{figure}
\vspace{0.0cm}
\hspace{0.0cm}
\psfig{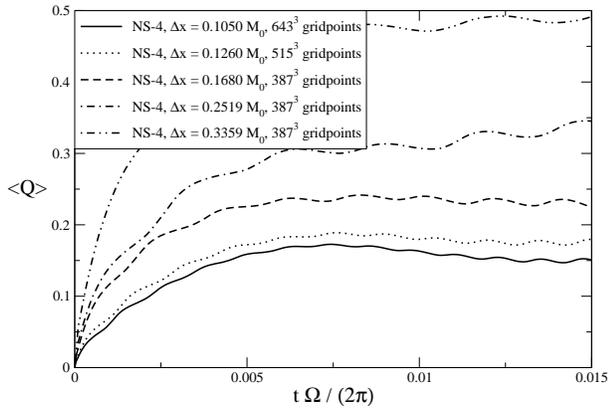}
\caption{
We plot the quantity $\langle Q \rangle$ (Eq.~\ref{eq:weight_q}) as a function
of time
for fully consistent general relativistic numerical
calculations, using CFQE configuration NS-4 as initial data
(see Table~\ref{tab:shortconfig}).  A variety of discretization
parameters $\Delta x$ are used.
}
\vspace{0.5cm}
\label{fig:q_ns4}
\end{figure}

\begin{figure}
\vspace{0.0cm}
\hspace{0.0cm}
\psfig{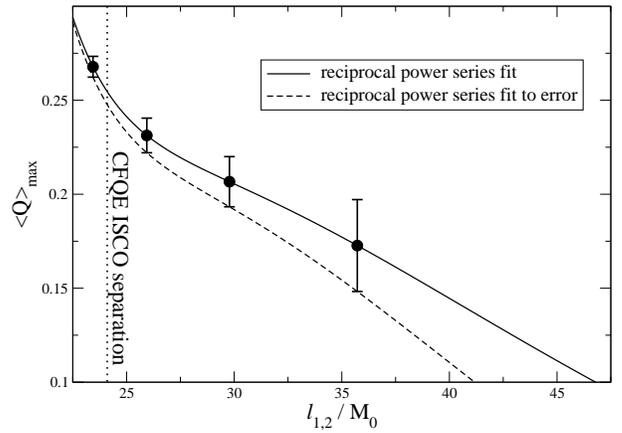}
\caption{
The maximum value of $\langle Q \rangle$ (Eq.~\ref{eq:weight_q}) obtained in 
our fully consistent general relativistic simulations using different
initial CFQE configurations with 
initial separations $\ell_{1,2}$.  The error bars are computed using
the error function Eq.~\ref{eq:error2}.
}
\vspace{0.5cm}
\label{fig:maxq}
\end{figure}

In Figures~\ref{fig:q_ns2},~\ref{fig:q_ns3},~and~\ref{fig:q_ns4}, we
plot $\langle Q \rangle$ as a function 
of time for increasing neutron star binary separations.
Notice that, for each separation, the shape of the curves appear to
be converging to something that is qualitatively similar
to that of Figure~\ref{fig:q_ns1}:  a quick   
increase to a maximum value.  Also note that this maximum value is, 
as expected, decreasing for increasing neutron star separation.   The best 
resolution for the maximum initial neutron star separation (NS-4)
in Figure~\ref{fig:q_ns4} (the solid line) has $\langle Q \rangle$ attaining a
maximum value that is already lower than $0.2$.  A natural question
then arises:  can one predict how far the initial separation of the
neutron stars should be in order that the maximum value of $\langle Q \rangle$
obtained in a short time scale in the consistent general relativistic
theory be bounded by some number, say, $\langle Q \rangle = 0.1$?  
One problem is 
in the details of the numerical simulations:  as the neutron star
separation increases, the computational resources demanded by the
problem become larger.  In other words, for a given fixed amount
of computational resources,  the ability to resolve each
neutron star (e.g., the number of discrete gridpoints across each
star) is diminished as the initial separation of the neutron stars
is increased.  This can be seen directly when comparing the 
resolutions used for the simulations performed for 
Figures~\ref{fig:q_ns1},~\ref{fig:q_ns2},~\ref{fig:q_ns3},~and~\ref{fig:q_ns4};
as the initial neutron star separation is increased in going 
from CFQE configuration NS-1 to NS-4, the resolution 
necessarily decreases.  These effects can be seen directly in 
the size of the error bars of 
Figure~\ref{fig:maxq}, where the maximum value of $\langle Q \rangle$ obtained 
in the short term general relativistic simulations is plotted
as a function of the initial geodesic separation $\ell_{1,2}$.  For each
data point, we use the maximum value
attained by the highest resolution curve (solid line) in each of 
Figures~\ref{fig:q_ns1}~-~\ref{fig:q_ns4}.  As always, it is important
to estimate the truncation and boundary errors in any numerical calculation.
Here, we use an error estimate of the form
\begin{eqnarray}
(\langle Q \rangle_{max})_{numerical} & = & 
(\langle Q \rangle_{max})_{exact} +  \nonumber \\
   & & C_1 (\Delta x) +
C_2 {(\Delta x)}^2 + \frac {C_3} {{r_b}^2}.
\label{eq:error2}
\end{eqnarray}
Note that we now include a term that is linear in the discretization
parameter $\Delta x$.   This is due to the fact that, although we are
using second order methods for the discretization of the Einstein
equations, the HRSC methods used for solving 
the relativistic hydrodynamics equations 
(described in Section~\ref{sec:discrete}) are only first order (in space)
accurate at points where the hydrodynamical variables obtain a local
extrema.  Also, while we expect the outer boundary condition
on general dynamical simulations to be better represented
by a $1/r_b$ error term, we note that in these short time scale 
simulations, the neutron stars are not even causally connected to the 
outer boundary.   The only boundary error in the calculation is that
due to the initial data CFQE configuration solve, whose 
boundary error decreases as $1/{r_b}^2$.  The error bars used in
Figure~\ref{fig:maxq} are computed using the 4 numerical
results obtained by varying the resolution and boundary placement
for each configuration, and solving
Eq.~\ref{eq:error2} for the constants $(\langle Q \rangle_{max})_{exact}$, 
$C_1$, $C_2$, and $C_3$.  The size of the error bar is then set 
equal to the largest of the absolute value of each individual error
term in Eq.~\ref{eq:error2}.  
With this generous estimate of the error of the maximum 
value of $\langle Q \rangle$, we can put a lower bound on the 
initial geodesic separation of 
the CFQE configuration that must be 
used as initial data in fully 
general relativistic simulations such that the Killing vector field
assumption is valid, e.g., $\langle Q \rangle \: < 0.1$.  While the value of $0.1$ may
be somewhat arbitrary (one may desire an even more stringent criterion),
the method we use to analyze the CFQE data is quite general.  In 
Figure~\ref{fig:maxq}, we fit a reciprocal power law,
$\frac {a_1}{{(\ell_{1,2})}} + \frac {a_2}{{(\ell_{1,2})}^2} +
\frac {a_3}{{(\ell_{1,2})}^3} + \frac {a_4}{{(\ell_{1,2})}^4}$,
to both the maximum value obtained in our fully consistent
general relativistic calculations, as well as to the lower
bound of the estimated error in our calculation.  We can see that 
one would have to use a CFQE configuration initial data set with
a geodesic separation between the neutron stars of 
at least $\ell_{1,2} = 46.8 \: M_0$ in order for the subsequent solution to the 
full Einstein field equations to satisfy the Killing field
assumption of the CFQE-sequence approximation to 1 part in 10
(this separation parameter could actually be as low
as $\ell_{1,2} = 41.2 M_0$, taking into account the errors
of our calculations, see Figure~\ref{fig:maxq}).  This separation corresponds
to roughly twice that of the ISCO separation.

\subsection{The conformal flatness assumption of the CFQE-sequence
approximation}
\label{sec:conflat}

\begin{figure}
\vspace{0.0cm}
\hspace{0.0cm}
\psfig{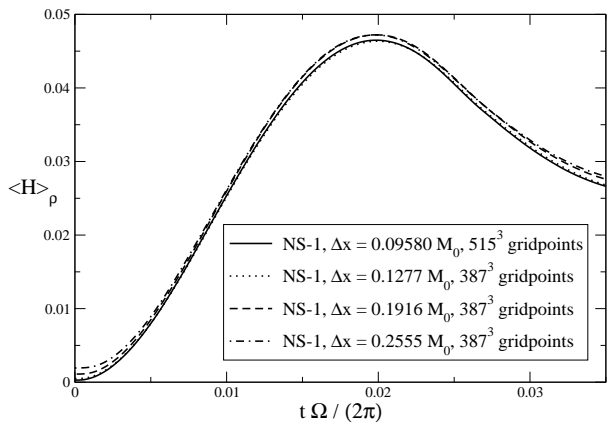}
\caption{
We plot the quantity $\langle H \rangle_\rho$ (Eq.~\ref{eq:weight_h_rho}) as a function
of time
for fully consistent general relativistic numerical
calculations, using CFQE configuration NS-1 as initial data
(see Table~\ref{tab:shortconfig}).  A variety of discretization
parameters $\Delta x$ are used.
}
\vspace{0.5cm}
\label{fig:hrho_zap1}
\end{figure}

One other
assumption in the CFQE-sequence approximation is
that of conformal flatness (CF).  It is often argued that this assumption
is, in some sense,
equivalent to assuming that there is no gravitational radiation in the
configuration.  However, the statement that
conformally flat configurations have zero gravitational 
radiation content is very questionable, especially in the case of the 
CFQE-sequence approximation which is not even
consistent with the full set of Einstein equations.  Here, we 
analyze this CF assumption in full general relativity.  We start
with a CFQE configuration as initial data, and perform fully consistent
general relativistic numerical evolutions, monitoring 
the conformal flatness of the spatial slices.  As we are using
the same slicing condition (maximal slicing) as that of the
CFQE-sequence approximation, we only require a 3-invariant that
will allow us to monitor the conformal flatness assumption
during the simulations in a coordinate
independent way.  The 3-Bach tensor is one such 3-invariant.  It
is defined on the spatial slice as
\begin{equation}
B_{ijk} = 2 {\cal D}_{[i} \left ( {}^{(3)}\!R_{j]k} - 
   \frac {1}{4} {\gamma}_{j]k} {}^{(3)}\!R \right ),
\end{equation}
and can be shown to vanish if and only if the 3-metric 
$\gamma_{ij}$ is conformally flat.  The Cotton-York tensor,
$H_{ij}$, is related to the 3-Bach tensor by
\begin{equation}
H_{ij} = {{\epsilon}^{mn}}_j B_{mni},
\end{equation}
where ${\epsilon}_{ijk}$ is the natural volume element 3-form. 
We define the scalar $H$ as the matrix norm of the Cotton-York
tensor, normalized by the size of the covariant derivative 
of the 3-Ricci tensor:
\begin{equation}
H = \frac {\left | H_{mn} \right |}
          {\sqrt{{\cal D}_i {}^{(3)}\!R_{jk} 
                 {\cal D}^i {}^{(3)}\!R^{jk}}},
\end{equation}
where, just as in the previous section, ${\left | H_{ij} \right |}$ denotes
the matrix norm of the components of $H_{ij}$ in our Cartesian coordinates.
Note that $H$ vanishes on conformally flat spatial slices, and is normalized
to provide a local measure for determining how much the spatial slice 
is deviating from conformal flatness.  
For a global measure, we define
the baryonic density weighted norm denoted
as $\langle H \rangle_\rho$, 
\begin{equation}
\langle H \rangle_\rho = 
   \frac {\int d^3\!x \: \left | H \right | \sqrt{\gamma} \rho W }
   {\int d^3\!x \: \sqrt{\gamma} \rho W },
\label{eq:weight_h_rho}
\end{equation}
where the integrals are taken to be over the entire spatial slice. 


In Figure~\ref{fig:hrho_zap1}, we analyze the CF assumption
of the CFQE-sequence approximation by plotting $\langle H \rangle_\rho$ as a function 
of time for fully consistent general relativistic simulations.  We use
for initial data the CFQE configuration 
NS-1 (see Table~\ref{tab:shortconfig}).  As it is always imperative to run
a numerical code at multiple resolutions and boundary placements to
assess the numerical errors, we
use a variety of discretizations and grid sizes.  Using this measure
of the violation of the conformal flatness assumption in the CFQE-sequence
approximation, we see that the assumption holds to roughly 1 part
in 20, for this initial data.  We can also see that this
measure, as compared to the measure for the QE assumption (Eq.~\ref{eq:weight_q}, Figure~\ref{fig:q_ns1}),
is not as sensitive to resolution.  In other words, the numerical
truncation error for this 
particular measure is not as large.

\begin{figure}
\vspace{0.0cm}
\hspace{0.0cm}
\psfig{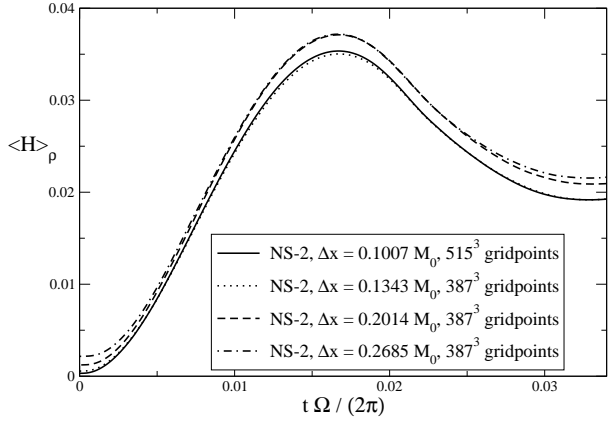}
\caption{
We plot the quantity $\langle H \rangle_\rho$ (Eq.~\ref{eq:weight_h_rho}) as a function
of time
for fully consistent general relativistic numerical
calculations, using CFQE configuration NS-2 as initial data
(see Table~\ref{tab:shortconfig}).  A variety of discretization
parameters $\Delta x$ are used.
}
\vspace{0.5cm}
\label{fig:hrho_zap2}
\end{figure}

\begin{figure}
\vspace{0.0cm}
\hspace{0.0cm}
\psfig{figure=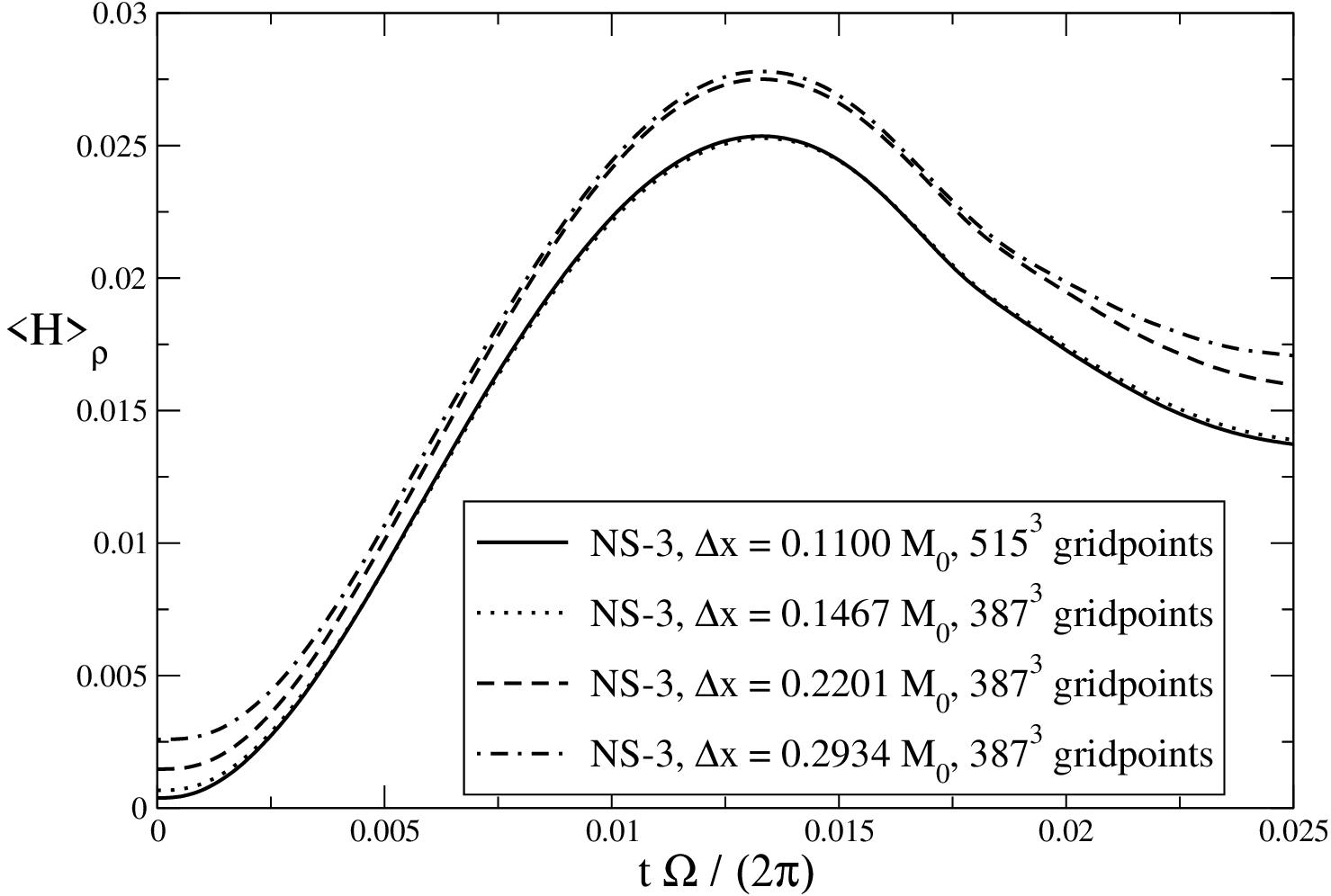,width=8cm}
\caption{
We plot the quantity 
$\langle H \rangle_\rho$ (Eq.~\ref{eq:weight_h_rho}) as a function
of time
for fully consistent general relativistic numerical
calculations, using CFQE configuration NS-3 as initial data
(see Table~\ref{tab:shortconfig}).  A variety of discretization
parameters $\Delta x$ are used.
}
\vspace{0.5cm}
\label{fig:hrho_zap3}
\end{figure}

\begin{figure}
\vspace{0.0cm}
\hspace{0.0cm}
\psfig{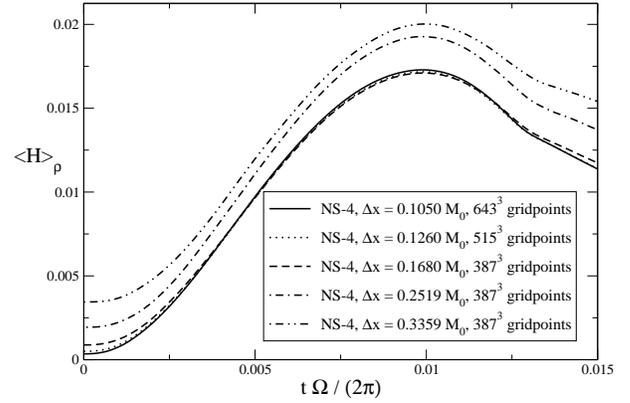}
\caption{
We plot the quantity 
$\langle H \rangle_\rho$ (Eq.~\ref{eq:weight_h_rho}) as a function
of time
for fully consistent general relativistic numerical
calculations, using CFQE configuration NS-4 as initial data
(see Table~\ref{tab:shortconfig}).  A variety of discretization
parameters $\Delta x$ are used.
}
\vspace{0.5cm}
\label{fig:hrho_zap4}
\end{figure}

\begin{figure}
\vspace{0.0cm}
\hspace{0.0cm}
\psfig{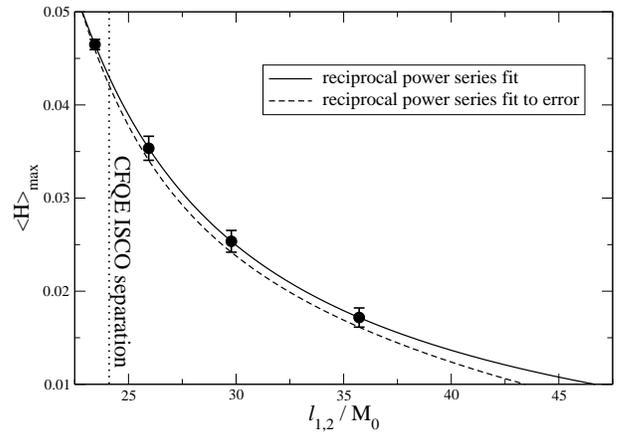}
\caption{
The maximum value of 
$\langle H \rangle_\rho$ (Eq.~\ref{eq:weight_h_rho}) obtained in
our fully consistent general relativistic simulations using different
initial CFQE configurations with
initial separations $\ell_{1,2}$.  The error bars are computed using
the error function Eq.~\ref{eq:error2}.
}
\vspace{0.5cm}
\label{fig:hrho_max}
\end{figure}

Again, we would expect that the CF assumption to be better for larger initial neutron
star separation.  In 
Figures~\ref{fig:hrho_zap2},~\ref{fig:hrho_zap3},~and~\ref{fig:hrho_zap4},
we plot the measure of the violation of the conformal flatness 
assumption $\langle H \rangle_\rho$, 
Eq.~\ref{eq:weight_h_rho}, in our 
general relativistic simulations using CFQE 
configurations NS-2, NS-3, and NS-4, 
respectively, as initial data (see Table~\ref{tab:shortconfig}).
We can see that the violation of the conformal flatness assumption
does, in fact, decrease with increasing initial neutron star separation.
As with the QE assumption, we can use 
the results of Figures~\ref{fig:hrho_zap1}~-~\ref{fig:hrho_zap4} to
predict the initial neutron star separation one would need for
a CFQE configuration to satisfy the CF assumption to some prescribed tolerance.  
For example, we may want to start our general
relativistic calculations with initial data
that corresponds to a CFQE configuration such that the 
error in the CF assumption is below one part in 100, as measured by the
quantity $\langle H \rangle_\rho$ (Eq.~\ref{eq:weight_h_rho}).  In 
Figure~\ref{fig:hrho_max}, we plot the maximum value of
quantity $\langle H \rangle_\rho$ as a function of initial geodesic separation
$\ell_{1,2}$ attained in our fully consistent general
relativistic numerical simulations using the four CFQE configurations
from Table~\ref{tab:shortconfig} as initial data
(see Figures~\ref{fig:hrho_zap1}~-~\ref{fig:hrho_zap4}).  Again, 
we use Eq.~\ref{eq:error2} and the method described in
Section~\ref{sec:killing} to compute the
numerical errors (both truncation errors and boundary
errors) made in the calculation.  We fit an inverse power 
series function
$\frac {a_1}{{(\ell_{1,2})}} + \frac {a_2}{{(\ell_{1,2})}^2} +
\frac {a_3}{{(\ell_{1,2})}^3} + \frac {a_4}{{(\ell_{1,2})}^4}$
to the four data points, as well as to the lower bound
of the error, in Figure~\ref{fig:hrho_max}.
As can be seen, one would have to
use initial data corresponding to a CFQE configuration with 
neutron star geodesic separation of approximately $46.7 M_0$ or greater in
order for $\langle H \rangle_\rho$ to be $0.01$ or less in
the subsequent solution
to the Einstein field equations coupled
to the hydrodynamics equations.  Recall from the previous section
that this separation would also satisfy the Killing field assumption
of the CFQE-sequence approximation to $10\%$.

\section{Long-term General Relativistic 
Numerical Simulations}
\label{sec:long_evolve}

In the previous section, we performed many short timescale (sub-orbital) 
general relativistic simulations using CFQE configurations as initial
data.  There, the focus was on determining the intrinsic error 
in using the CFQE configurations as initial data to 
model astrophysical neutron star binaries.  Here, we study the 
long timescale (i.e., multiple orbital timescale) evolution of the
system in order to investigate
the stability and accuracy of long term fully general relativistic 
numerical integrations of the neutron star binary system.

There are several reasons why it is extremely 
difficult to study
large separation binary neutron stars on time
scales longer than the orbital period in full numerical 
relativity in an accurate fashion.
One basic reason is apparent from the dynamics of
two Newtonian point masses in circular motion.
The orbital period $T_{orb}$ increases as the separation $D_{sep}$ of 
the masses as
\begin{equation}
T_{orb} \sim {\left ( D_{sep} \right )}^{\frac {3}{2}}.
\end{equation}
Thus, any simulation of neutron stars with larger separation will
naturally have to be run to longer times in order to capture
a single orbital period.  Also, from the standard quadrapole
formula, the energy in gravitational waves $E_g$ emitted per
unit time for particles in circular motion decreases with
increasing separation $D_{sep}$ as
\begin{equation}
\frac {dE_{g}}{dt} 
\sim {\left ( D_{sep} \right )}^{-5}.
\end{equation}
Obviously, the time for any simulation to track the orbital motion
of compact binaries through the plunge phase to the final merger
will be extremely sensitive to the initial separation.

We can get some idea of the
computational resources needed to accurately simulate
binary neutron stars with initial geodesic proper separations of between
$25 \: M_0$ and $35 \: M_0$ (which corresponds to roughly
$2.5 \: R_{NS}$ and $4.5 \: R_{NS}$, where $R_{NS}$ is the neutron star
radius) from the results of the previous section.  In 
Section~\ref{sec:short_evolve}, we performed 
general relativistic numerical calculations using roughly 
these separations (see Table~\ref{tab:shortconfig}) for only 
several percent of one orbital period.   It took numerical
configurations of over $500^3$ gridpoints in order to obtain 
resolutions high enough for a confident prediction of the
error of the simulations.  
In order to perform simulations on orbital timescales,
we need to increase the simulation times by two orders of magnitude.  
While this is already quite difficult, the situation is even more demanding: 
such a simulation time is much greater than the light
crossing time of our computational domain.  The neutron stars will
no longer be causally disconnected from our dynamical boundary 
conditions as was the case in the short timescale simulations
performed in the previous section.  The computational boundary
will therefore have a much greater affect on the simulation 
results if put at the 
same spatial location.  One may need to greatly increase the
distance from the center of mass of the system to the computational
boundary, but one must be careful not to sacrifice the spatial resolution
at the same time.

Preliminary results obtained in~\cite{Shibata01a} suggest that 
waveforms calculated from a numerical relativity binary inspiral 
simulation can be done in an accurate fashion (e.g., with
errors less than 10\%) only when the extraction radius is
approximately one gravitational wavelength 
$\lambda_{g}$ away from 
the center of mass of the system (for comparison, the NS-4 calculations 
from the previous section has the outer boundary at only $0.08 \lambda_{gw}$). 
However, gravitational wave extraction is just one aspect of 
our simulation requirements.  
Other effects including spacetime dynamics and dynamics of the 
binary may require the boundary to be even farther away.  
These facts, coupled with the complexities involved in
solving the full Einstein field equations by computer, render the
problem of obtaining simulations accurate enough to probe the
details of large-separation orbiting binary neutron stars a
most difficult one.

\begin{table}
\begin{tabular}{|c|c|c|c|c|}  \hline \hline
\hspace{0.0cm} { Configuration} \hspace{0.0cm}  &
\hspace{0.0cm} { grid size} \hspace{0.0cm}  &
\hspace{0.0cm} { $\Delta x / M_0$} \hspace{0.0cm}  &
\hspace{0.0cm} { $r_b / \lambda_{gw}$} \hspace{0.0cm} &
\hspace{0.0cm} { $r_{id} / \lambda_{gw}$ } \hspace{0.0cm} \\ \hline \hline
{NS-A} &
   { 643x643x325} &
   { 0.2085} &
   { 0.257} &
   { 0.257} \\ \hline
{NS-B} &
   { 323x323x165} &
   { 0.2085} &
   { 0.129} &
   { 0.257} \\ \hline
{NS-C} &
   { 313x313x160} &
   { 0.2607} &
   { 0.156} &
   { 0.160} \\ \hline
{NS-D} &
   { 259x259x133} &
   { 0.2607} &
   { 0.129} &
   { 0.160} \\ \hline
{NS-E} &
   { 163x163x85} &
   { 0.4171} &
   { 0.129} &
   { 0.160} \\ \hline \hline
\end{tabular}
\vspace{0.0mm}
\caption{The computational domain configurations used for the large 
timescale binary neutron star general relativistic 
simulations.  All large timescale simulations are performed
with a CFQE configuration characterized by an orbital angular 
velocity of $\Omega M_0 = 0.01204$, where the 
geodesic separation of the neutron stars is $\ell_{1,2} = 27.57 M_0$.
The gravitational wavelength $\lambda_{gw}$ corresponding to this
configuration is 
$\lambda_{gw} = \frac {1}{2} (\frac {2 \pi}{\Omega}) = 260.9 M_0$.
$r_b$ denotes the (coordinate) distance from the center of the orbiting binary
to the boundary of the computational domain. $r_{id}$ denotes
the (coordinate) distance between the center of the orbiting binary
to the boundary of the computational domain used in solving
for the CFQE initial data configuration.
}
\vspace{0.5cm}
\label{tab:longconfig}
\end{table}

In this section we analyze the numerical evolutions of one
particular 
CFQE initial configuration which has a larger
geodesic separation $\ell_{1,2}$ than that of the CFQE ISCO
configuration.  Specifically, the angular orbital velocity of
the CFQE configuration we use exclusively in this section is
$\Omega M_0 = 0.01204$.  This configuration has a
geodesic separation of $\ell_{1,2} = 27.57 M_0$, and
corresponds to the second smallest $\Omega$ data point shown in
Figure~\ref{fig:cfqe_richard}.
According to the study in section~\ref{sec:short_evolve}, this 
configuration has a violation of the QE and CF assumptions at 
the $22\%$ and $3\%$ levels, respectively, soon after the evolution starts.
We numerically evolve this
CFQE initial data configuration using our fully consistent general relativistic
treatment.  The gauge conditions used for these simulations are the
``1+log'' equation for the lapse $\alpha$ (Eq.~\ref{eq:1pluslog}) and
Eq.~\ref{eq:shiftcondition} for the shift vector $\beta^i$.  
There is no need to use maximal slicing in this section as 
comparing to the CFQE-sequence is no longer the point.

In Table~\ref{tab:longconfig}, we list the properties of the various 
computational domains,
varying both the resolution and outer boundary placement, used for our 
long timescale numerical evolutions.  Our numerical implementation 
allows us to use a different location for the outer boundary
of the computational domain for our initial data solve of
the CFQE configuration as that used for the dynamical evolution.
We denote $r_{id}$ as the shortest coordinate distance between the
center of our computational domain and the computational boundary of
our cubical domain used in solving the initial data problem for
the CFQE configuration.  We denote $r_b$ as the shortest coordinate
distance between the center of our computational domain and
the computational boundary of our cubical domain used in our
fully consistent general relativistic numerical simulations.
Note that in Table~\ref{tab:longconfig}, the number of grid
points refers to that used in the full dynamical evolution.  For
those computational domains where $r_{id} > r_b$, we have used the 
same resolution $\Delta x$ with a larger number of grid points to
solve for the CFQE initial data configuration.

\begin{figure}
\vspace{0.0cm}
\hspace{0.0cm}
\psfig{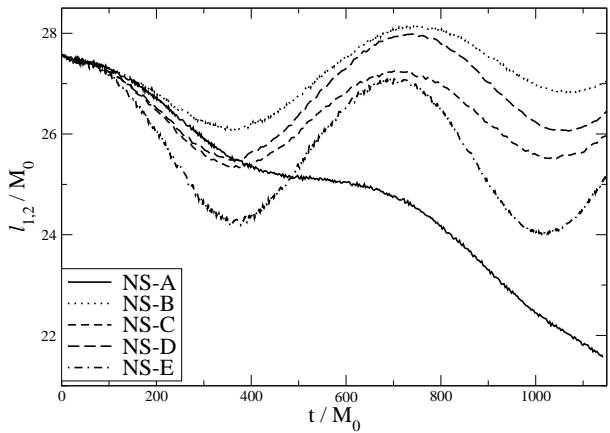}
\caption{
We plot the evolution of the geodesic distance between the maximum
rest mass density of the two neutron stars as a function of
coordinate time for various resolutions and outer boundary placements
(see Table~\ref{tab:longconfig} for configuration specifications).
}
\vspace{0.5cm}
\label{fig:geodist}
\end{figure}

In Figure~\ref{fig:geodist}, we plot the geodesic separation
$\ell_{1,2}$ as a function of coordinate time $t$ using this
initial data for the various computational domains listed 
in Table~\ref{tab:longconfig}.  Qualitatively, we observe a number
of interesting features.  First of all, note that all of the simulations in
Figure~\ref{fig:geodist} 
display an eccentricity in the
orbit. 
Some level of eccentricity is expected due to the fact that the
CFQE configuration is constructed with the explicit assumption that the
time derivative of the separation $\ell_{1,2}$ exactly vanishes,
as argued in~\cite{Mora02,Miller03a}.  That is,
the assumption of the radial velocity of
the binaries to be exactly zero instantaneously (which is
done by assuming the existence of a timelike helical
Killing vector field in the construction of CFQE configurations) is not
consistent with the astrophysically relevant
scenario of quasicircular binary evolution, where the magnitude of the
radial velocity, while small, never vanishes.
One can obtain some
idea regarding the level of orbital eccentricity intrinsic
to CFQE configurations by analyzing the
point-particle dynamics
in the post-Newtonian approximation~\cite{Miller03a}.
In~\cite{Miller03a}, it is shown in the context
of the post${}^{9/2}$-Newtonian approximation that if the
assumption of circular motion is used to construct the initial conditions
(that is, the conditions $\dot{r} = 0$ and $\ddot{r} = 0$ are used
to specify the initial conditions), then the resulting orbit
will have a nonzero eccentricity whose value depends on the
initial separation.  

Ideally, we would like to directly compare our general relativistic
results for the eccentricity of the orbits apparent in
Figure~\ref{fig:geodist} with the post-Newtonian results
in~\cite{Miller03a}.  The reason we are unable to do this
at the present time is obvious from Figure~\ref{fig:geodist}.
Note the large difference between the eccentricity for computational
domain configuration NS-A as compared to the other configurations.  In
fact, it is obvious
from Figure~\ref{fig:geodist} that the NS-A configuration
is qualitatively different from the other configurations.   Recall that
all configurations NS-A through NS-E use the same CFQE configuration
as initial data;  note from Table~\ref{tab:longconfig} that
the only parameters that differ in producing the 
results from Figure~\ref{fig:geodist} are the spatial resolution of the 
discretized computational domain, the location of the outer boundary 
of the computational domain during numerical evolution where the
dynamical boundary conditions are imposed, and the location of 
the outer boundary of the computational domain used in solving for the CFQE
initial data configuration.  A cursory study of 
Table~\ref{tab:longconfig} reveals the major difference between
computational domain configuration NS-A and the others, namely, the 
location of the outer boundary $r_b$ of 
the computational domain during numerical
evolution.   The boundary distance $r_b$ of the computational 
domain configuration NS-A is slightly larger than
$1/4$ of the gravitational wavelength $\lambda_{gw}$ characterized 
by the CFQE initial data configuration,
while all of the other configurations have 
$r_b \leq 0.156 \: \lambda_{gw}$.  

We must always quantify the errors of any numerical
calculation.
Here, our numerical errors originate from two distinct sources.  The
first source of error is the truncation
error associated with our discretization parameter $\Delta x$, while
the second source of error is
induced by the outer boundary conditions imposed on our 
numerical calculation.  While the behavior of the truncation error
is well understood (it is a local error which scales as an integer
power of the discretization parameter $\Delta x$), the assessment of
the effect of the boundary conditions is not as straightforward.
Theoretically, one could
place the outer boundary of the computational domain 
sufficiently far away so that 
the outer boundary condition would not be causally 
connected to the compact objects, and thus would have no effect 
on them during numerical evolution.  However, this is not a practical
solution, due to the limitations of computational 
resources (especially for a unigrid code;  adaptive mesh refinement could
be used in this direction).
As our outer boundary {\it is}
causally connected to the neutron stars in our simulations, we must 
attempt to assess the errors introduced by the outer boundary
conditions in our numerical simulations.  We assume that the
error induced by the outer boundary conditions can be expanded 
in terms of powers of $1/r_b$, and that the error goes to $0$ as
$r_b \rightarrow 0$.  We therefore assume an error function for
the eccentricity $e$ as
\begin{equation}
e_n = e_{exact} + C_1 (\Delta x) + C_2 {(\Delta x)}^2 +
   \frac {C_3}{r_b} + \frac {C_4}{{r_b}^2}.
\label{eq:eccentricerror}
\end{equation}
where $e_n$ denotes the measured value of eccentricity from our numerical
solution using discretization parameter $\Delta x$ and outer boundary
location $r_b$.  Using the definition of eccentricity
defined in~\cite{Miller03a} in which the eccentricity of the
orbit is calculated from the orbital separation as a function
of time, we compute the eccentricity associated with each 
simulation shown in Figure~\ref{fig:geodist}.  We find that
${(e_n)}_{NS-A} = 0.0124$,
${(e_n)}_{NS-B} = 0.0327$,
${(e_n)}_{NS-C} = 0.0397$,
${(e_n)}_{NS-D} = 0.0434$, and
${(e_n)}_{NS-E} = 0.0605$.
We can then solve for the unknown quantities $e_{exact}$, $C_1$, $C_2$,
$C_3$, and $C_4$ in Eq.~\ref{eq:eccentricerror}.
We find that the Richardson extrapolated value of
the eccentricity is $e_{exact} = -0.127$, and that the leading
error terms for the truncation error and boundary error are
$C_1 \Delta x = 0.067$ and $C_3/r_b = 0.11$, respectively.  
Above and beyond the fact that the Richardson extrapolated
value of the eccentricity $e_{exact}$ is negative, the obvious sign that
we are not in the convergence regime (i.e. that the higher order
terms neglected in the error expansion of Eq.~\ref{eq:eccentricerror}
are not relatively small) is that $|e_{exact} - e_n|$ is larger than
the error terms in Eq.~\ref{eq:eccentricerror}.  As the computational
resources available at the present time
do not currently allow us to use our unigrid code to simultaneously
decrease
the discretization parameter $\Delta x$ further and increase the
distance $r_b$ from the center of mass to the outer boundary,
we must admit that we can, at this time, make no definite conclusion 
as to the inherent eccentricity in CFQE configurations used as
initial data in numerical relativity.  However, the prospect of 
being able to determine this point in the near future is good;
we can expect both mesh refinement techniques and better outer
boundary conditions to greatly aid in reducing the 
errors induced by the outer boundary in our numerical calculations.

\subsection{orbital decay rate}

\begin{figure}
\vspace{0.0cm}
\hspace{0.0cm}
\psfig{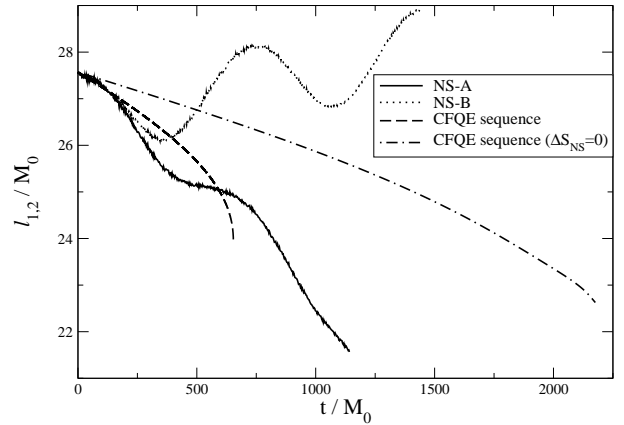}
\caption{
The geodesic separation $\ell_{1,2}$ as a function of time.
Shown are the results from our fully consistent general relativistic
calculation NS-A and NS-B (see Table~\ref{tab:longconfig}).  
Also shown are predictions from the CFQE-sequence approximation, i.e.
the curve obtained by numerically integrating
Eq.~\ref{eq:dldt}.  The curve labeled ``CFQE sequence'' was
constructed using the standard definition of
binding energy, Eq.~\ref{eq:bindingenergy} 
(see Figure~\ref{fig:cfqe_richard}), and is terminated at the 
CFQE ISCO point ($\ell_{1,2} / M_0 = 24.0$).  The curve labeled
``CFQE sequence $(\Delta S_{NS}=0)$'' was constructed using
the binding energy $E_b^\prime$, Eq.~\ref{eq:newbindingenergy},
where the neutron star spin remains constant throughout the 
entire CFQE sequence (see Figure~\ref{fig:cfqe_richard_new} and 
the discussion in Section~\ref{sec:nsspin}), and is terminated at
the neutron star touching point ($\ell_{1,2} / M_0 = 22.6$).
}
\vspace{0.0cm}
\label{fig:geodist_cfqe}
\end{figure}

Recall from 
Sections~\ref{sec:cfqesequence}~and~\ref{sec:nsspin} 
that the binding energies $E_b$ and
$E_b^\prime$
(Eqs.~\ref{eq:bindingenergy}~and~\ref{eq:newbindingenergy},
respectively) shown in 
Figures~\ref{fig:cfqe_richard}~and~\ref{fig:cfqe_richard_new} 
represent the binding energy
of each neutron star binary 
CFQE configuration as a function of geodesic separation $\ell_{1,2}$
(the orbital angular velocity $\Omega$ is monotonically increasing
with decreasing geodesic separation $\ell_{1,2}$).
In the CFQE-sequence approximation, this binding energy is slowly
converted to gravitational wave energy, and 
Figures~\ref{fig:cfqe_richard}~and~\ref{fig:cfqe_richard_new}
tell us how much 
gravitational radiation energy is produced for changes in 
geodesic separation.  We can approximate the rate of energy loss 
at any specific point in this sequence using the standard 
quadrupole formula (see, e.g.,~\cite{Wald84}), which reduces to
\begin{equation}
\frac {dE_{gw}}{dt} = \frac {128}{5} M^2 R^4 {\Omega}^6
\label{eq:dedt}
\end{equation}
for two point particles of mass $M$ in circular orbit with 
radius $R$ and orbital angular velocity $\Omega$.  We interpolate
the data represented in 
Figures~\ref{fig:cfqe_richard}~and~\ref{fig:cfqe_richard_new}
with a cubic spline to obtain the effective binding energies
$E_b$ (Eq.~\ref{eq:bindingenergy}) and $E_b^\prime$ 
(Eq.~\ref{eq:newbindingenergy}) as a function of geodesic
separation $\ell_{1,2}$.  We can then easily find 
$dE_b/d\ell_{1,2}$ as a function of geodesic separation.
An estimate of the time rate of change of geodesic
separations is then
\begin{equation}
\frac {d\ell_{1,2}}{dt} = 
   \frac { dE_{gw}/dt}{dE_b/d{\ell_{1,2}}},
\label{eq:dldt}
\end{equation}
which can be numerically integrated to produce the geodesic 
separation $\ell_{1,2}$ as a function of time predicted 
from the CFQE-sequence approximation.  We plot these
functions in Figure~\ref{fig:geodist_cfqe}, along with 
results from our fully consistent general relativistic calculation
NS-A and NS-B.

Note that the standard CFQE-sequence prediction of the evolution
of the geodesic separation $\ell_{1,2}$, labeled
``CFQE sequence'' in Figure~\ref{fig:geodist_cfqe}, terminates
at a neutron star geodesic separation 
$\ell_{1,2} = 24.0 \: M_0$, whereas the modified 
CFQE-sequence prediction, labeled
``CFQE sequence $(\Delta S_{NS}=0)$'', terminates at a  smaller
geodesic
separation of
$\ell_{1,2} = 22.6 \: M_0$.  This is due to the fact that
the standard effective binding energy $E_b$ 
defined by Eq.~\ref{eq:bindingenergy}
actually attains a minimum before the neutron stars touch
(see Figure~\ref{fig:cfqe_richard}), whereas the effective
binding energy for the CFQE-sequence where the spin of the
neutron stars do not change ($E_b^\prime$ defined by
Eq.~\ref{eq:newbindingenergy}) is monotonically decreasing
as the neutron star separation decreases, right through to
the point where the neutron stars are touching 
(see Figure~\ref{fig:cfqe_richard_new}).  Therefore,
the first ``CFQE-sequence'' prediction in Figure~\ref{fig:geodist_cfqe}
terminates at this minimum point (defined as the ISCO configuration
of the sequence), whereas the second
``CFQE sequence $(\Delta S_{NS}=0)$'' prediction
in the figure terminates at the CFQE configuration where the neutron
stars are touching.  One might expect that a full solution
to the Einstein equations using a CFQE configuration as initial data
might, in fact, have an evolution of the geodesic separation
of the neutron star binaries $\ell_{1,2}$ that would lie somewhere in between
the two CFQE-sequence approximations shown in Figure~\ref{fig:geodist_cfqe}.
After all, these two CFQE-sequence approximations represent the 
extreme cases of the evolution of the neutron star spin; the first
represents complete tidal locking during the entire 
evolution, the second represents absolutely
no change in the spin state of the individual neutron stars during
the entire evolution.  Of course,
the actual solution to the full Einstein equations would be expected to 
couple, to some amount, the increasing orbital angular frequency of
the binary neutron stars to
the spin of the individual stars.  The magnitude of this coupling would
determine which of the two CFQE-sequence approximations
shown in Figure~\ref{fig:geodist_cfqe} would be considered to
be ``more'' correct.  It is exactly this question which could be
answered with our fully consistent general relativistic calculations.
However, in Figure~\ref{fig:geodist_cfqe} we once again see the dramatic
effect of the outer boundary conditions on our numerical simulations.
Note that computational domain configurations used in the calculations
NS-A and NS-B have the same spatial
resolution $\Delta x = 0.2085 \: M_0$ and outer boundary location 
$r_{id} = 0.257 \: \lambda_{gw}$ used for solving the CFQE initial data
configuration.
The only difference
between the configurations NS-A and NS-B is the outer boundary
placement used during the dynamical evolution:  
the outer boundary of the computational
domain used during numerical simulation NS-A is twice as far
away from the center of mass as that used in simulation NS-B.
One may be tempted to conclude that
the calculation NS-A could be
said to be ``more correct'' than NS-B, since the outer boundary is located
farther away for the NS-A configuration.  However, one 
must be careful when trying to apply physical intuition to
numerical results that are not in the convergence regime. In this
case, for instance,
there is no reason to believe
that the error induced by the outer boundary on our calculation
is a monotonic function of $r_b$ (whereas, e.g., it is true in 
general that the truncation error is monotonic in $\Delta x$ as
$\Delta x \rightarrow 0$).  Due to the wave nature of the
gravitational radiation being emitted by the binary
neutron stars, the boundary errors could have
an oscillating component.  This makes it even more difficult to 
try to do a Richardson extrapolation type of error analysis in realistic
compact object simulations in numerical relativity.

While it should be expected that the extraction of 
gravitational radiation for a numerically generated spacetime could be
highly sensitive to the location of the outer boundary
(see~\cite{Shibata01a}), there have been, up to now, no results showing
what effect the outer boundary conditions can have on the details of the
orbits of compact binaries on timescales larger than one orbital period.
Here, we see that the dynamical outer 
boundary conditions can, and do, significantly
affect the orbital parameters of compact binaries during numerical
evolution.  

\section{Conclusions}
\label{sec:conclusions}

To date, the only fully general relativistic simulations of 
corotating binary, $\Gamma = 2$ polytropes is~\cite{Shibata99d}.  
The study in~\cite{Shibata99d} differs from the present study in
several important ways.  One basic difference is that while we focus on 
the capability of simulating astrophysical realistic neutron 
star binaries, \cite{Shibata99d}
focuses on the dynamics of the final merger of the two neutron stars.  Thus,
the initial data used in~\cite{Shibata99d} are CFQE 
configurations either at ISCO separation, or
closer than ISCO separation 
(ISCO separation here means simply the neutron star separation of
the unique
configuration that corresponds to a minimum of the 
binding energy in the constant rest mass CFQE-sequence,
e.g., in Figure~\ref{fig:cfqe_richard}).
Also, in the study~\cite{Shibata99d}, it was found necessary to manually 
decrease the orbital angular momentum of each CFQE configuration
at the level of several percent,
in order to precipitate the binary merger more quickly.  
Complementary to that approach, we are trying to use numerical 
relativity as a tool to assess the fidelity of the CFQE-sequence
approximation itself.  We found that in order to perform simulations 
of the neutron star binary systems compatible with realistic
astrophysical scenarios, one must perform
simulations using initial data at a distance considerably 
larger than the ISCO separation when using corotating CFQE configurations
as initial data.

We have outlined a generic method for analyzing the regime of
validity of the CFQE-sequence approximation and have applied this
method to the case of equal mass, corotating
binary neutron stars.  We have found that, for corotating neutron stars,
the violation of the timelike helical Killing vector field existence
assumption was an order of magnitude larger than the violation of the
assumption of conformal flatness.  Specifically, we have demonstrated
that initial data specified by 
a CFQE configuration with neutron stars having an initial geodesic
separation of less than $47 \: M_0$ 
(which is slightly more than $6$ neutron star radii,
or roughly twice the ISCO configuration separation) would produce
a solution to the Einstein field equations that violates the 
Killing vector field assumption by more than 
$1$ part in $10$;  the conformal flatness
assumption would be violated by more than $1$ part in $100$.
We thus conclude that, in the corotating case,
the CFQE-sequence approximation for neutron star separations of $47 \: M_0$
and less violates the Einstein field equations at a level larger
than $10\%$, and thus
numerical simulations starting with similar CFQE configurations as initial
data cannot,
therefore, be considered as approximating a realistic neutron star
binary inspiral.

We note that the violations of the assumptions of the CFQE-sequence
approximation that we observe in our general
relativistic calculations for the corotating binary systems 
occur on timescales that are two orders
of magnitude shorter than the orbital timescale.  We suspect that this 
may be due to interactions between the spin assumption 
of the individual neutron stars (the corotating assumption)
and the CFQE assumptions.  We have shown that the 
characteristic shape of the effective binding
energy curve within the CFQE approximation is highly sensitive to 
the spin kinetic energy of the individual neutron stars.  We have shown
that if we subtract out the spin kinetic energy of the neutron stars
in the construction of the effective binding energy (which approximates the
case where the neutron star spin does not increase as the orbital angular
velocity increases),
then the resulting binding
energy curve will have no minimum, and thus the CFQE-sequence
approximation would not predict the existence of an ISCO configuration.
We speculate that specifying 
neutron stars with irrotational spin states in the CFQE-sequence
approximation may yield a smaller violation of the Einstein field
equations for a fixed neutron star separation.  
The analysis we have developed in this paper can be used for a 
detailed investigation of this effect.  More interestingly, the 
analysis we have developed might provide a way to determine a spin 
state most consistent with the CFQE approximations, and hence 
provide a more realistic set of initial data that can be used to 
start simulations at a smaller initial separations. 

We have shown that, for our specific neutron star models, we require
a resolution of approximately $\Delta x =  0.1 M_0$ in order to 
adequately resolve the neutron stars (``adequately resolve'' here
refers to verifying that we are in the convergence regime through
an appropriate Richardson extrapolation technique; this is a much more
stringent condition than has been typically used in numerical 
relativity studies 
to date involving neutron stars and black holes).
This resolution scale is over three orders of 
magnitude smaller than
the characteristic wavelength of the gravitational
radiation emitted during the last five to ten orbits of the neutron 
star inspiral process.   

We have also shown that the location of the outer boundary 
of the computational domain can have a significant impact on the 
details of the evolution of the compact objects on timescales
of the orbital period.  Specifically, we have seen that changing 
the linear dimensions of our computational domain from $0.3 \: \lambda_{gw}$
to $0.5 \: \lambda_{gw}$ can significantly impact the dynamics of 
binary neutron stars during the first several orbits.
This should serve as a warning to the numerical 
relativity community studying simulations of compact binaries with the
hope of extracting gravitational wave information: not only
will the outer boundary inhibit the actual process of extracting
the gravitational waves, but they also directly affect the 
sources of the gravitational waves themselves.  While our
dynamical boundary conditions are not the best choice, and a more
consistent treatment, e.g. constraint preserving boundary
conditions, would most likely improve the situation, it may be that
numerical relativists will be forced to push the outer
boundary of the computational domain to the 
``local wave zone'' (which in this case means $r_b \geq \lambda_{gw}$)
in order to provide realistic gravitational 
waveforms suitable for use as templates
in gravitational wave detectors.

Unfortunately, this makes the numerical study of 
orbiting compact objects in numerical relativity particularly hard.
Note that every time we increase
the resolution in our 3D simulations by a factor of two, keeping
the outer boundary $r_b$ fixed, we must use 16 times the amount 
of computational resources to perform any particular simulation.  
Also, every time we increase the 
outer boundary distance $r_b$ by a factor of two, we must use
8 times the amount of computational resources.   Therefore, if we
wanted to simultaneously increase both the resolution 
and the outer boundary distance by a factor of 2, we would require over
two orders of magnitude more computational resources.  While we have shown
in this paper that it is possible to track the details of
finite sized compact objects in full numerical relativity, what remains
is to be able to do so in such a way that all of the numerical
errors (both truncation and boundary errors) can be demonstrated to
be small over a timescale of several orbital periods.  This will be
an extremely challenging task, given the current level of computational
resources available.  It may be necessary to employ 
mesh refinement in order
to accurately simulate all of the physical degrees of freedom that
we are interested in.

\section{Acknowledgement}
\label{ack}

It is a pleasure to thank
Abhay Ashtekar, Comer Duncan, David Garfinkle, 
Lee Lindblom, David Meier, Peter Miller, Thierry Mora, 
Masaru Shibata, Kip Thorne,
and Clifford Will
for useful discussions and comments.  
We also thank Nikolaos Stergioulas for providing us with
code for calculating the ADM mass of
stationary, uniformly rotating polytropic stars.  Our application code
which solves the Einstein equations coupled to the relativistic hydrodynamic
equations, along with our various multigrid elliptic solvers, 
uses the Cactus Computational Toolkit~\cite{Cactusweb} for
parallelization and high performance I/O.

Financial support for this research has been
provided by the ASC project (NSF Phy 99-79985) and the 
Jet Propulsion Laboratory (account 100581-A.C.02) under contract with the 
National Aeronautics and Space Administration.
Computational resource support has been provided by the 
NSF NRAC projects MCA02N022 and MCA93S025, and the NAS at Ames, NASA.



\end{document}